\documentclass[twocolumn]{aastex62}
\usepackage{amsmath}
\usepackage{footmisc}
\makeatletter
\newcommand*{\rom}[1]{\expandafter\@slowromancap\romannumeral #1@}
\makeatother
\graphicspath{{./}{figures/}}

\received{March 31, 2018}
\revised{April 8, 2018}
\accepted{\today}
\submitjournal{ApJ}

\shorttitle{Disentangling Blended \emph{K2} Photometry}
\shortauthors{Payne et al.}

\begin{document}
\title{Disentangling Blended \emph{K2} Photometry: Determining the Planetary Host Star \footnote{Accepted Astronomical Journal 14th, September 2018}}

\correspondingauthor{Alan Payne}
\email{alan.payne@usq.edu.au}

\author[0000-0002-0786-7307]{Alan N. Payne}
\affil{University of Southern Queensland \\
Faculty of Health, Engineering and Sciences \\
Toowoomba Qld 4350, Australia}
\nocollaboration

\author{David R. Ciardi}
\affiliation{NASA Exoplanet Science Institute/Caltech \\
Pasadena, CA 91125, USA}
\affiliation{University of Southern Queensland \\
Faculty of Health, Engineering and Sciences \\
Toowoomba Qld 4350, Australia}
\nocollaboration

\author{Stephen R. Kane}
\affiliation{University of California Riverside \\
Department of Earth Sciences \\
Riverside, CA 92521, USA}
\nocollaboration

\author{Brad Carter}
\affiliation{University of Southern Queensland \\
Faculty of Health, Engineering and Sciences \\
Toowoomba Qld 4350, Australia}
\nocollaboration



\begin{abstract}
The presence of companion stars, whether bound or unbound, make correct identification of the planetary hosting star difficult when a planet has been detected through a photometrically blended transiting event. We present an approach that uses a combination of light curve analysis and stellar modeling to disentangle 8 \emph{K2} photometrically blended binary systems that have either a confirmed or suspected planet to identify the probable host star. The key to our approach is the use of the mean stellar density, calculated using the transit geometry, as a discriminator. The approach is strongly dependent on the difference in magnitude between the stars and the observed transit depth, which is corrected by the flux ratio between the two stars. While our approach does not lead to a definitive result for all systems, we were able to determine for the 8 systems in this case study: two systems where the primary was highly likely to be the planet hosting star ($>$ 90\% likelihood); three systems where the primary was likely to be the hosting star ($>$ 55\% likelihood); one system where the secondary was likely to be the planet hosting star ($>$ 55\% likelihood); and two systems where we were uncertain which star was the planet hosting star ($\approx$ 50\% likelihood to be either the primary or the secondary). Two systems, denoted by their EPIC number, EPIC 201546283 and EPIC 201862715, had confirmed planets, K2-27b and K2-94b respectively, with radii of $\mathrm{4.0 \pm 0.1}$ and $\mathrm{16.4 \pm 2.9}$ $\mathrm{R_\oplus}$ respectively.\\

\end{abstract}



\section{Introduction} \label{sec:intro}

The primary focus of the \emph{Kepler} \citep{2010Sci...327..977B} and the reinvented \emph{K2} \citep{2014PASP..126..398H} missions are to detect and characterize planets by capturing high signal to noise time series photometry of planetary transit of host stars. The \emph{Kepler} spacecraft has observed hundreds of thousands of stars, which includes numerous multi-star systems. The exact number of multi-star systems within the \emph{Kepler} data set is uncertain, but it has been proposed that ~30\%--40\% of \emph{Kepler} host stars (stars that host one or more planets) are also multi-star systems (\cite{2012AJ....144...42A}; \cite{2013AJ....146....9A}; \cite{2014AJ....148...78D}; \cite{2015ApJ...805...16C}; \cite{2017AJ....153...71F}). Likewise, \emph{K2} host stars have a similar fraction of stars with stellar companions as \emph{Kepler} and solar-type field stars (\cite{2018AAS...23110902M}). This fraction appears to be similar to the fraction of solar neighborhood stars with stellar companions (approximately 46\%)  (\cite{2010ApJS..190....1R}). 

The ability of the \emph{Kepler} detector to resolve multi-star systems is limited due to it having a relatively large pixel size (approximately $4\arcsec$ on sky\footnote{Characteristics of the Kepler space telescope: https://keplerscience.arc.nasa.gov/the-kepler-space-telescope.html}). The consequence of this design means that the flux from unknown stellar objects can fall within the same \emph{Kepler} pixel resulting in confounding photometry, known as a blend. If the observed flux from a pixel does contain two or more stars, then the flux levels for the primary star (the brighter star in the system) will be inflated by the other stars in the system. Where a transiting planet has also been detected, the increased flux levels dilutes the transit depth leading to an underestimation of the planets radius \citep{2015ApJ...805...16C}. When combined with a planetary mass, underestimating a planet's radius directly impacts the derived bulk density (\cite{2015ApJ...805...16C}; \cite{2015ApJ...801...41R}; \cite{2017AJ....154...66F};  \cite{2017AJ....153..117H}). In addition, if a planetary transit has been revealed in the photometry, there will be a high level of uncertainty surrounding which star is the planetary host star.   

Determining which star in a \emph{Kepler}/\emph{K2} photometrically blended signal is the planetary host star requires follow-up ground based photometry of each star in the system. This can be problematic if the stars causing the contamination are close on the sky, and/or aren't bright enough to be resolved, and/or there is a lack of certainty around the stellar parameters for each star in the system. Its worth noting that whilst its reasonable to assume that stars separated by $1\arcsec$--$2\arcsec$ on the sky will be gravitationally bound \citep{2017AJ....153..117H}, non-bound systems (either foreground or background stars) will also present the same distortion to the observed flux on the \emph{Kepler}/\emph{K2} detector. We therefore use the terminology \lq multi-star system' to denote that there are two or more stellar objects falling on one \emph{Kepler}/\emph{K2} pixel -- the term is not meant to infer that the system is necessarily bound. High-resolution imaging by ground based telescopes has been used to identify blended systems at a resolution of $\geq 0.02\arcsec$. If high-resolution imaging indicates that the \emph{Kepler}/\emph{K2} photometry is blended and a transiting planet has been identified then the question still remains -- which star in the multi-star system hosts the planet and what is the true radius of the planet? 

Estimating the true planetary radii from an observed diluted transit depth is heavily dependent on the difference in magnitude between the primary star and the remaining stars in the system and which star in the system the planet orbits \citep{2015ApJ...805...16C}. For example, if a planet is in orbit around the primary star and there is a large difference in the magnitude between the host and the companion/s then the planet's radius will be slightly underestimated. Similarly, if a planet is in orbit around the secondary star and the difference in magnitude is large, then the estimation of the planet's radius will be significantly underestimated. 

In this paper, we will investigate an approach to determine which star in a multi-star system, whether bound or unbound (all systems are assumed to be unbound), is the planetary host star, thereby allowing us to correct reported estimations of planetary radii. For a small selection of \emph{K2} planetary hosting stars \citep{2016ApJS..226....7C} we have laid out an approach to distinguish which star the planet orbits. A few attempts have been made to determine which star a planet orbits in multi-star systems for example, \emph{Kepler-13}; \citep{2017AJ....154..158B}, \emph{Kepler-296}; \citep{2015ApJ...809....7B} and \emph{K2-136}; \citep{2018AJ....155...10C}. \cite{2018AJ....155...10C} used the stellar density as a discriminator for which star hosts a Neptune-sized planet in a binary system in the Hyades Cluster; \emph{K2-136}. Our approach will be to build upon the approach by \cite{2018AJ....155...10C} with the aim of publishing new radii for planets located in multi-star systems selected from \cite{2016ApJS..226....7C}. We begin by outlining our target selection in Section 2, present our approach in Section 3 and our main results in Section 4 and 5. We discuss our results and conclude in Section 6.

\section{Target Selection} \label{sec:targets}

To undertake our case study we started with the 48 multi-star systems identified in \cite{2016ApJS..226....7C}. They form part of a subset of 196 systems (FGK type stars and late type M and K type stars) discovered from the first year of the \emph{K2} mission (Campaigns 0--4). These multi-star systems were identified using follow-up high resolution ground based imaging using bandpass filters I and/or J and/or K. \cite{2016ApJS..226....7C} presents a full description of their target selection. 

From the 48 multi-star systems we selected 8 multi-star systems as our final target systems. These systems are treated as if they are visual binaries and were initially selected if they had at least one confirmed, or a possible planetary companion, and excluded false positives; 2 targets have confirmed planetary companions and 6 have possible planetary companions (see Table \ref{targets}). The planetary candidate associated with one of our targets, EPIC 205148699, may be a smaller stellar object \citep{2016ApJS..226....7C}, but we still include this target in our analysis as it doesn't affect our outcomes. In addition, the selected systems had the following stellar and planetary parameters and their uncertainties (data obtained from \cite{2016ApJS..226....7C} and the ExoFOP Archive \footnote{\label{exofop}ExoFOP Archive: https://exofop.ipac.caltech.edu/k2/}): stellar surface gravity ($\log g$); metallicity ([Fe/H]); effective temperature ($T_\mathrm{eff}$); and the planets period. These parameters enabled us to utilize ExoFAST \citep{2013PASP..125...83E}, a transit modeling algorithm based on Mandel and Agol's \citep{2002ApJ...580L.171M} transit fitting algorithm. Our targets are tabulated in Table \ref{targets}.

\begin{deluxetable*}{cccccccccccc}[ht!]
  \tabletypesize{\scriptsize}
  \tablecolumns{11}
  \tablecaption{\label{targets} Selected \emph{K2} Targets}
  \tablehead{
    \colhead{EPIC ID} & \colhead{RA} & \colhead{Dec} &
   \colhead{Period} &
    \colhead{Period} & \colhead{$T_\mathrm{eff}$} & \colhead{$T_\mathrm{eff}$} & \colhead{$\log g$} & 
    \colhead{$\log g$} & \colhead{{[}Fe/H{]}} & \colhead{{[}Fe/H{]}} &\colhead{Status}  \\
    \colhead{System} & \colhead{} & \colhead{} &
    \colhead{} & \colhead{Uncertainty} & \colhead{} & \colhead{Uncertainty} & \colhead{} & 
    \colhead{Uncertainty} & \colhead{} & \colhead{Uncertainty} &\colhead{}  \\ \hline
    \colhead{} & \colhead{} & \colhead{} &
    \colhead{(days)} &
    \colhead{(days)} & \colhead{(K)} & \colhead{(K)} & \colhead{} & \colhead{} & \colhead{} & \colhead{}\\ \hline
    \colhead{(1)} & \colhead{(2)} & \colhead{(3)} &
    \colhead{(4)} & \colhead{(5)} & \colhead{(6)} &
    \colhead{(7)} & \colhead{(8)} & \colhead{(9)} & \colhead{(10)} & \colhead{(11)} & \colhead{(12)}    
  }
  \startdata  
201546283  &	11:26:03.64  &	01:13:50.66           &	6.771315  &	0.000079  &	5343  &	161  &	4.494  &	0.07  &	-0.026  &	0.15  & Confirmed\\
201862715  &	11:43:38.01  &	06:33:49.41           &	2.6556777  &	0.00000044  &	5920  &	54  &	4.519  &	0.007  &	0.1  &	0.1 & Confirmed \\
205148699  &	16:52:38.51  &	-19:09:41.94          &	4.377326  &	0.000045  &	5928  &	144  &	4.081  &	0.342  &	-0.171  &	0.2 & Candidate\textsuperscript{a}\\
206011496  &	22:48:07.57  &	-14:29:40.88          &	2.369193  &	0.000087  &	5622  &	132  &	3.868  &	0.585  &	-0.475  &	0.4 & Candidate\\
206061524  &	22:20:13.77  &	-13:06:52.66          &	5.87975  &	0.0008  &	4056  &	81  &	4.883  &	0.05  &	-0.367  &	0.27 & Candidate\\
206192335  &	22:10:39.4   &	-09:53:23.14          &	3.59912  &	0.00024  &	5620  &	137  &	4.478  &	0.055  &	-0.179  &	0.24 & Candidate\\
210958990  &	04:11:00.96  &	22:19:31.22           &	1.702301  &	0.000018  &	6116  &	259  &	4.243  &	0.15  &	-0.101  &	0.21 & Candidate\\
211147528  &	03:58:35.32  &	25:23:18.92	    &  2.349455 & 	0.0000092 &	7056	 &	457  &	4.064  &    0.165 &    -0.016  &  0.1 & Candidate\\
  \enddata
  \tablecomments{Tabular summary of the selected KOI (multi-star systems). Column (1) presents the \emph{Kepler} identification number. Columns (2) and (3) describe the position of the KOI. Columns (4) and (5) provide the planetary period and its associated uncertainty. Columns (8)--(11) provide stellar parameters, effective temperature ($T_{eff}$), surface gravity (log$g$), and metallicity ([Fe/H]) and their associated uncertainties. Column (12) indicates if the planet is confirmed or a candidate. The a superscript for planetary candidate associated with EPIC 205148699 may be stellar in nature.}  
 \end{deluxetable*}
\section{The Approach} \label{sec:appr}
\label{ta}

Our approach in determining which star is the planet hosting star utilizes the relationship between the mean stellar density and the effective temperature. We found the mean stellar density for each star in our systems using the observed transit geometry as described in \cite{2003ApJ...585.1038S} and then compared our results to model values of the stellar density. How closely the mean stellar density reflects the modeled values, estimated from the transit parameter geometry, determines the likelihood that the planet is associated with that star. 

Section \ref{ta} is split into two subsections. First, we present our methodology for disentangling photometrically blended sources. This allows the determination of the apparent \emph{Kepler} bandpass magnitudes, $K_p$, for each contributing source in our binary systems and their flux ratios (the flux ratio is used as a scaling factor, which is applied to the transit depth to facilitate the determination of the mean stellar density for our target stars). Finally, we present our methodology for determining the effective temperature for the stars in our target systems.

\subsection{Determining Mean Stellar Density}
\label{kmfr}
\subsubsection{Disentangling Photometrically Blended Sources}
\label{KpandK}
\begin{figure*}[htp]
\includegraphics[width=20cm,height=15cm]{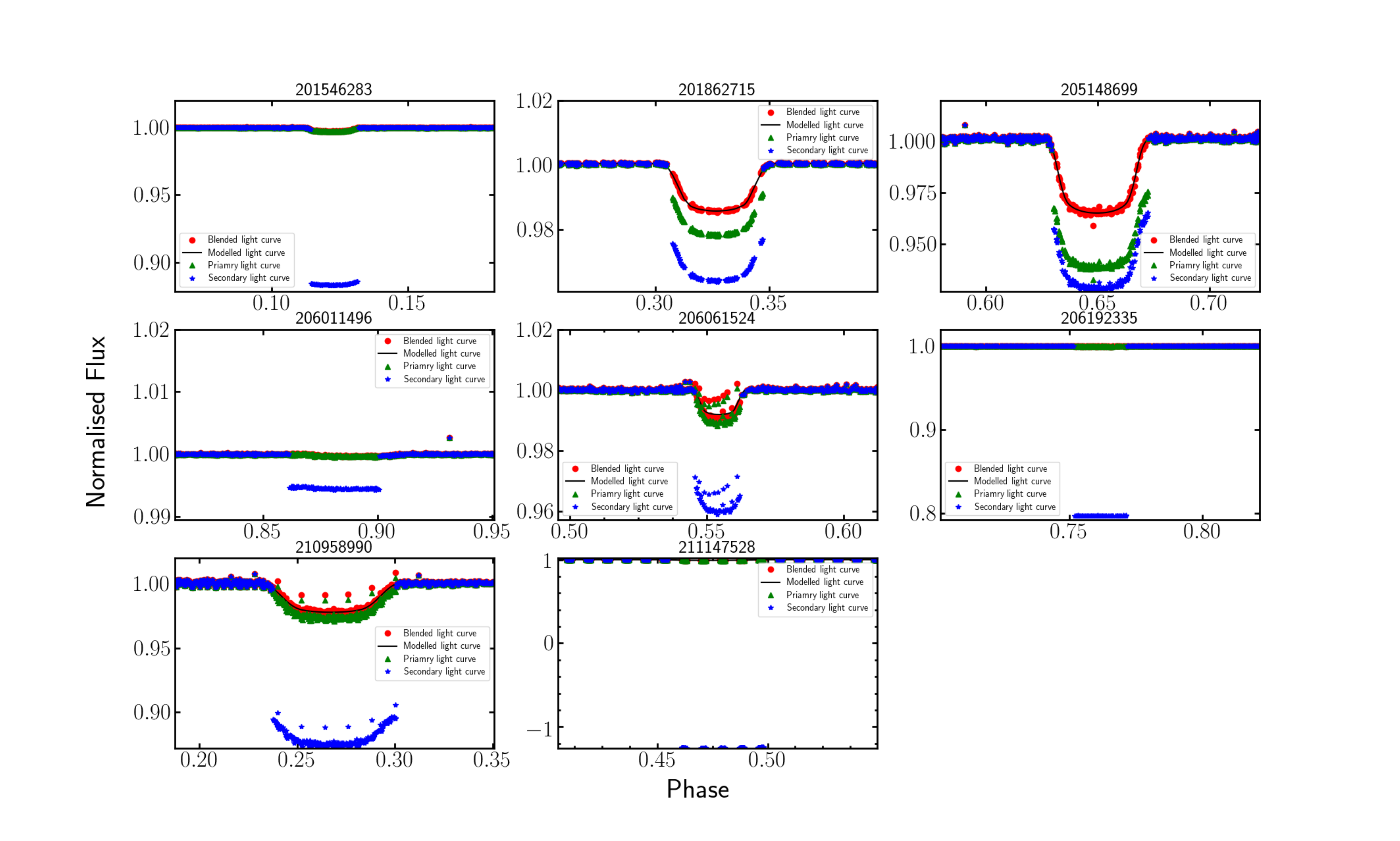}
\caption{Modeled light curve represented by the black line, for the observed phase folded transits, red dots for each of the eight target systems. Also shown, is the predicted phase folded transits after the flux has been separated making the assumption the planet is in orbit around the primary star, green triangles and second, assuming the planet is orbit around the secondary stellar object, blue stars. The effect on the transit depth is clearly visible. Also the last panel, EPIC 211147528, shows clear evidence to suggest that the planet hosting star is the primary because the normalized flux for the secondary is negative and therefore meaningless. \label{blends}}
\end{figure*}

The consequence of observing a photometrically blended signal with an embedded planetary transit is that it dilutes the transit depth and, since the mean stellar density is dependent on the transit depth, when calculated using transit parameters (see Equation \ref{sdf} ), it leads to spurious results for the mean stellar density. Figure \ref{blends} clearly shows the impact on the transit depth compared to the observed transit depth. It presents phase folded transits of all eight blended target systems after making an assumption the planet is hosted by the primary and likewise assuming the planet is hosted by the secondary. In addition, it isn't meaningful to find the mean stellar density for a blended source, its only meaningful to find the mean stellar density for each individual star within a system. 

The last panel in Figure \ref{blends} (EPIC 211147528) shows that the normalized flux is negative for the secondary star. This can not be the case because even if the planet was the same size or bigger then the star then under that scenario all the light would be blocked and the normalized flux would be reduced to zero, not negative. This is clear evidence to support that the planet hosting star for the EPIC 211147528 system is the primary. This was also discussed and shown by \cite{2016ApJS..226....7C}. Even though we know that primary star is the host star we still included this system in our list of target systems because it is a good test case to verify our approach at determining the planet host star (discussed later).             

To determine the relative photometric contributions from each star in our target systems, we measured the relationship between $K_p$ and the infrared passband magnitude, $K$ by construction a bivariate brightness distribution between $K_p$ and $K$ from a subset of all \emph{K2} stars. Using the magnitude difference in the $K$ passband data collected from high resolution follow-up imaging, we derived for both stars the deblended \emph{Kepler} bandpass magnitudes.

The high resolution $K$ follow-up imaging of our target systems was analyzed using the Aperture Photometry Tool (APT) \citep{2012PASP..124..737L}, which allowed the determination of $K$ passband magnitude differences between both stars in each of the target systems, $\Delta K$. From the measured $\Delta K$ and the 2MASS $K_s$ magnitudes we derived an apparent real $K$ magnitude for each star in our target systems using equations \ref{fluxr} and \ref{mag}. The flux ratio was determined using, 
\begin{figure*}[htp]
\includegraphics[width=20cm,height=15cm]{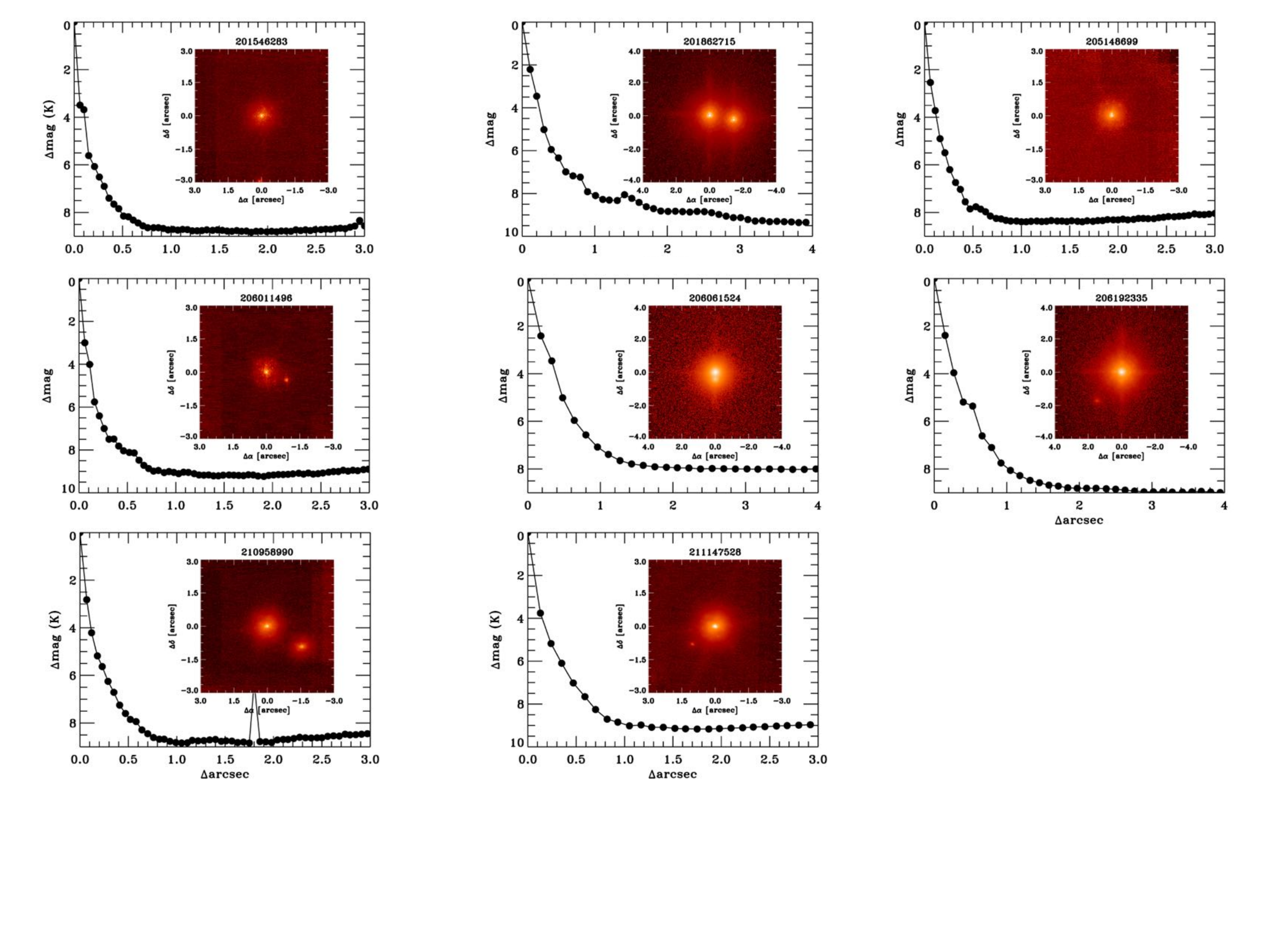}
\caption{Contrast sensitivity and inset images for our target systems using a K filter observed with either the Keck Observatory Keck II 10 meter adaptive optics system or Palomar Observatory 5 meter Hale Telescope adaptive optics system. The contamination from the secondary stellar object is clearly visible in all systems. The $5\sigma$ contrast limits for additional companions, in $\Delta m$, are plotted against angular separation in arcseconds. The black points represent one step in FWHM resolution of the image. Images sourced from the ExoFOP Archive\footref{exofop}\label{images}}
\end{figure*}
\begin{equation}
f_1 / f_2 = 10^{-(\Delta K / 2.5)},
\label{fluxr}
\end{equation}
where $f_1/f_2$ is the flux ratio. The real apparent $K$ magnitudes were determined using,   
\begin{subequations}
\begin{align}
m_1 &=  m_\mathrm{tot} - 2.5 \log \left[\frac{1} {(1 + f_2/f_1)} \right] \: \mathrm{and} \\
m_2 &= m_\mathrm{tot} - m_1,
\end{align}
\label{mag}
\end{subequations}
where $m_1$ is the magnitude of the primary, $m_2$ is the magnitude of the secondary and $m_\mathrm{tot}$ is the 2MASS $K_s$ magnitude. Figure \ref{images} shows the results of these calculations for all our systems.

With $K$ magnitudes for both the stars in our systems, we use this information to estimate a value for $K_p$ for both our stars using the relationship between $K_p$ and $K$. This relationship was constructed by using a sample of \emph{K2} stars from the EPIC catalog \citep{2016ApJS..224....2H}, and where possible multi-star systems were excluded. The initial sample contained a mix of approximately 250,000 dwarf and giant stars. Since our target systems are assumed to consist of only dwarf stars, we first separated the giants from the dwarfs to ensure that our sample was representative of our target systems. We adopted a similar approach to that demonstrated in \cite{2011AJ....141..108C}; the sample was split into giants and dwarfs using a surface gravity--effective temperature Hertzsprung--Russell diagram, via the following:
\begin{equation*}
\log(g) \geq \begin{cases}
3.6 &\text{if $T_{\mathrm{eff}} \geq 6500$} \\
4.0 &\text{if $T_{\mathrm{eff}} \leq 5000$} \\
5.3 - 2.6 \times 10^{-4} &\text{if $5000 < T_{\mathrm{eff}} <  6500$.}
\end{cases}
\end{equation*}

\begin{figure*}[ht!]
\plotone{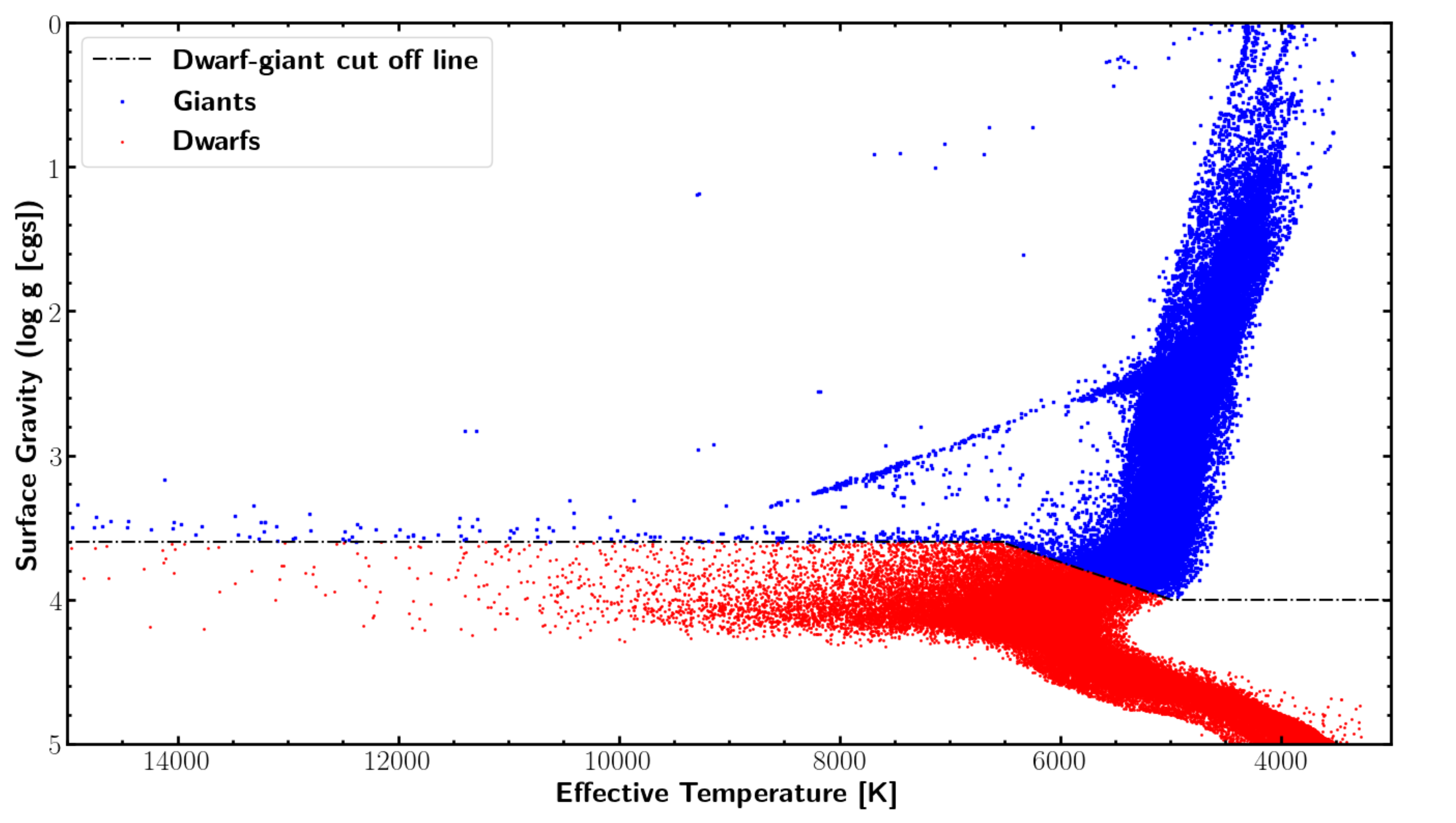}
\caption{The dashed line represents the separation between dwarfs and giants. The dwarfs are colored red and are located below the dashed line while the giants are blue and are located above the dashed line.}
\label{loggTeff}
\end{figure*}

The separation between dwarfs and giants and is shown in Figure \ref{loggTeff} by the dashed line with the dwarfs located below the dashed line while the giants are located above. The giant subset shows a non-physical systematic between effective temperatures of 5000K to 9000K and surface gravity of 2.5$g/cm^3$ to 3.5$g/cm^3$. The systematic is also seen in \cite{2016ApJS..224....2H}. However, since we are only focused on the dwarf subset this systematic will not effect our results.       

\begin{figure}[htp]
\plotone{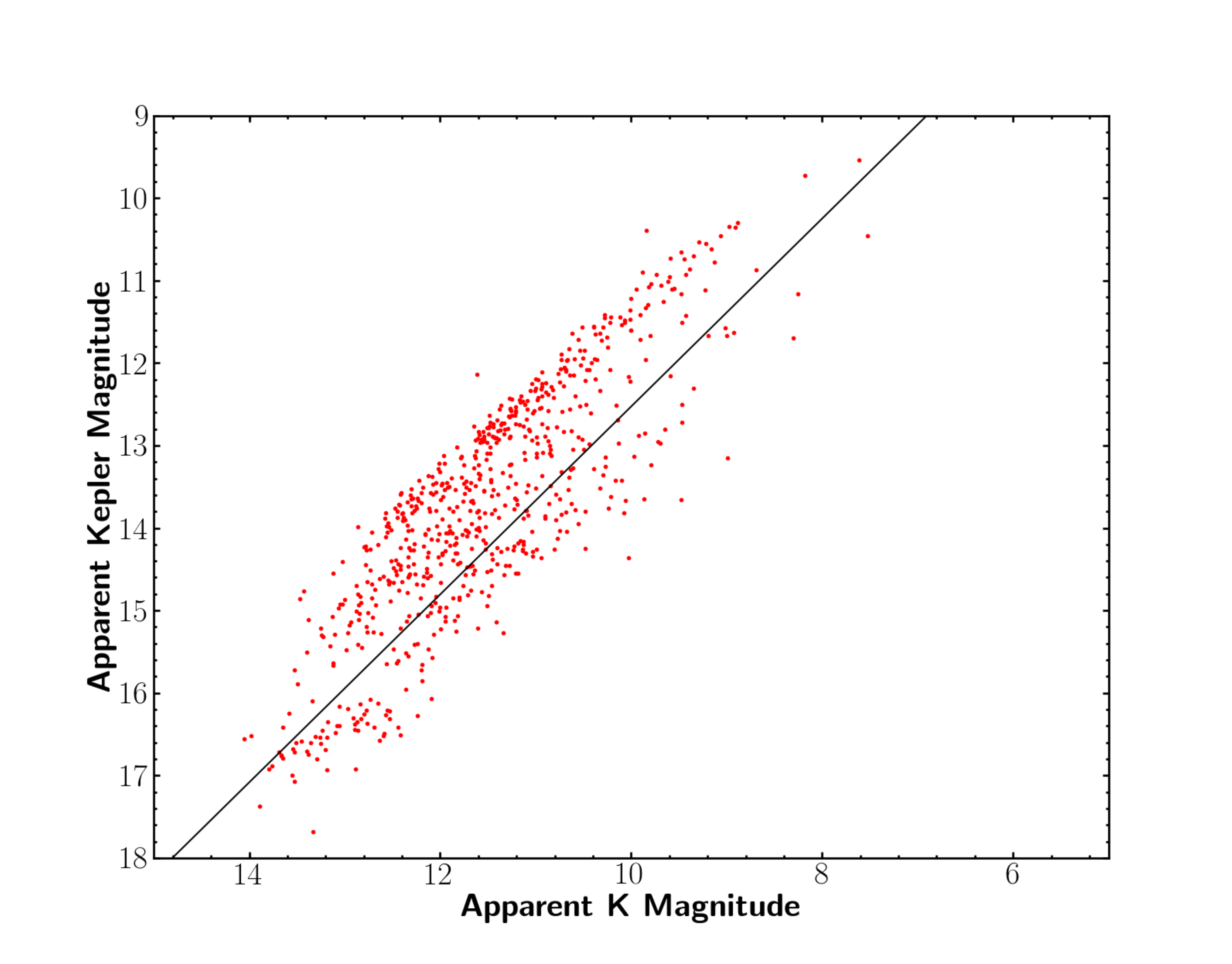}
\caption{Model of $K2$ dwarf stars showing the relationship between $K_p$ and $K$.}
\label{bivar}
\end{figure}

We further constrained our sample stars by selecting stars with effective temperatures between 3000K and 8000K and uncertainties less than 2\%, surface gravities with uncertainties less than 5\% and distances with uncertainties less than 10\%. We then modeled these data with a simple linear model to derive an expression that relates $K_p$ to $K$. The model coefficients and uncertainties, shown in the parenthesis, are presented in Equation \ref{linear}. The model is also depicted graphically in Figure \ref{bivar}. Using the derived $K$ magnitudes for our target component stars we were able to estimate $K_p$ magnitudes, 
\begin{equation}
\label{linear}
\begin{split}
K_p  = 1.12944(0.02889)K + 0.67721(0.33653), \\
\end{split}
\end{equation}
and uncertainties for each stellar component in the system. The results are tabulated in Table \ref{Kp_flux}.

\begin{deluxetable*}{ccccccccccc}[htp]
\tabletypesize{\scriptsize}
\tablecolumns{11}
\tablecaption{\label{Kp_flux} $K$, $K_p$ Magnitudes And Flux ratio}
\tablehead{
    \colhead{EPIC ID} & \colhead{$K_p$} & \colhead{$K_p$} & \colhead{$K_s$} & \colhead{$K_s$} & \colhead{$K_{p(p)}$} & \colhead{$K_{p(p)}$} & \colhead{$K_{p(s)}$} & \colhead{$K_{p(s)}$} & \colhead{f2/f1}  & \colhead{f2/f1} \\
    \colhead{System} & \colhead{} & \colhead{Uncertainty} & \colhead{} & \colhead{Uncertainty} & \colhead{} & \colhead{Uncertainty} & \colhead{} & \colhead{Uncertainty} & \colhead{} & \colhead{Uncertainty} \\ \hline
    \colhead{} & \colhead{(mag)} & \colhead{(mag)} & \colhead{(mag)} & \colhead{(mag)} & \colhead{(mag)} & \colhead{(mag)} & \colhead{(mag)} & \colhead{(mag)} & \colhead{} & \colhead{} \\ \hline
    \colhead{(1)} & \colhead{(2)} & \colhead{(3)} & \colhead{(4)} & \colhead{(5)}  & \colhead{(6)} & \colhead{(7)} & \colhead{(8)} & \colhead{(9)} & \colhead{(10)} & \colhead{(11)}    
  }
\startdata  
201546283  &	10.74  &	0.02  &	14.49  &	0.02  &	12.80  &	0.46  &	17.05  &	0.54  &	0.020   &	0.013\\
201862715  &	9.26  &	0.02  &	9.78  &	0.03  &	11.13  &	0.43  &	11.72  &	0.44  &	0.582  &	0.330\\
205148699  &	11.23  &	0.06  &	11.39  &	0.08  &	13.36  &	0.47  &	13.54  &	0.48  &	0.849  &	0.526\\
206011496  &	9.34  &	0.03  &	12.15  &	0.03  &	11.22  &	0.43  &	14.40  &	0.49  &	0.054  &	0.032\\
206061524  &	11.84  &	0.02  &	13.26  &	0.03  &	14.05  &	0.48  &	15.65  &	0.51  &	0.228  &	0.147\\
206192335  &	10.25  &	0.02  &	16.56  &	0.06  &	12.26  &	0.45  &	19.38  &	0.59  &	0.001  &	0.001\\
210958990  &	10.94  &	0.02  &	12.49  &	0.02  &	13.04  &	0.46  &	14.79  &	0.49  &	0.199  &	0.124\\
211147528  &	10.63  &	0.02  &	15.95  &	0.02  &	12.68  &	0.46  &	18.69  &	0.57  &	0.004  &	0.003\\
\enddata
\tablecomments{$\Delta K_p$ magnitude, flux ratio and uncertainties for our target stars. The (p) and (s) subscripts denotes the primary and secondary respectively. Column (1) is the \emph{K2} system identification number. Columns (2)--(5) presents $K$ and their uncertainties. Columns (6)--(9) presents $K_p$ and their uncertainties. Columns (10)--(11) is the $K_p$ flux ratio and their uncertainties.}  
\end{deluxetable*}

\subsubsection{Transit Modeling}
\label{transmod}

Before calculating the mean stellar density of our target stars, the period ($P$), the transit depth ($\Delta F$), the transit duration  ($t_{T}$), and the impact parameter ($b$) needed to be found. To ensure a self-consistent modeling of the transits across all systems, we modeled the 30 minute cadence \emph{K2} light curves using EXOFAST \citep{2013PASP..125...83E} with Markov Chain Monte Carlo analysis. The transit fit parameters are tabled in Table \ref{stellardensity} and are similar to the results presented in \cite{2016ApJS..226....7C}.
 
\subsubsection{Mean Stellar Density}
\label{sd}

Once the relative photometric contribution each star makes to an observed blended photometric signal and the transit parameters are known, the task of determining which star is the planetary hosting star may be determined. We used the mean stellar density as a discriminator to determine which star is the planetary hosting star. This was derived for each star in our target systems using the derived transit parameters, as outlined by \cite{2003ApJ...585.1038S}. The rational behind using the mean stellar density is based on the assumption that only one of the derived stellar densities can be right as the planet can not be hosted by both stars in the system unless the planet is in a circumbinary orbit. Our targets are not in circumbinary orbits because the periodicity of our target planets are short, indicating they lie close to their parent stars. If a star has an an observable planetary transit, \cite{2003ApJ...585.1038S} showed that the mean stellar density $\rho_{*}$ can be estimated directly from the transit parameters:
\begin{equation}
\begin{split}
\rho_{*} \approx \frac{3\pi}{P^{2}G} \bigg \{\frac{(1 + \sqrt{\Delta F_\mathrm{{obs}}})^{2} - b^2[1 - \sin ^{2}(t_{T}\pi/P)]} {\sin ^{2}(t_{T}\pi/P)}\bigg \}^{3/2},
\end{split}
\label{sdf}
\end{equation}
where $\Delta F_\mathrm{{obs}}$ is the observed transit depth. Equation \ref{sdf} is stated as an approximation because we are assuming that $\rho_{*} \gg \rho_\mathrm{{planet}}(\mathrm{Radius_{planet}}/\mathrm{Radius_{*}})^3$ \citep{2010arXiv1001.2010W}. We note here that this differs slightly to what was published in \cite{2003ApJ...585.1038S} as we believe that in \cite{2003ApJ...585.1038S} they used $\rho_* = M_*/R_*^3$ and not $\rho_* = M_*/(4/3)\pi R_*^3$ for the mean stellar density.

To determine the mean stellar density for both the primary and the secondary stars in our target systems, we first assumed that the planet is in orbit around the primary star and then likewise that it was in orbit around the secondary star. Both cases required us to use Equation \ref{sdf}, which depends on knowing the transit depth, transit duration, period and impact parameter. The only parameter that varies between the two cases is the transit depth. To find the correct transit depth for each case we applied a correction factor to the observed transit depth. The correction factor was derived by first deducing that the total observed flux is,  
\begin{equation}
F_\mathrm{obs} = f_1 + f_2,
\label{obsflux}
\end{equation}
and the transit depth for the primary was found using,
\begin{equation}
\Delta F_p = \bigg [\frac{f_1 + f_2}{f_1} \bigg ]\Delta F_\mathrm{obs} = \left(1 + \frac{f_2}{f_1} \right) \Delta F_\mathrm{obs},
\label{obsfluxp}
\end{equation}
where $\Delta F_p$ is the primary transit depth and the flux ratio, $f_2/f_1$, is the the flux ratio described in Section \ref{KpandK}. Similarly, making the assumption that the planet is in orbit around the secondary star, the transit depth was found using,
\begin{equation}
\Delta F_s = \bigg [\frac{f_1 + f_2}{f_2} \bigg ]\Delta F_\mathrm{obs} = \left(1 + \frac{f_1}{f_2} \right) \Delta F_\mathrm{obs},
\label{obsfluxs}
\end{equation}
where $\Delta F_s$ is the transit depth of the secondary star. Equations \ref{obsfluxp} and \ref{obsfluxs} can now be substituted into Equation \ref{sdf} to find an expression for the mean stellar density of both the primary and secondary stars in terms of the observed blended flux. If we assume the planet is hosted by the primary the mean stellar density becomes,
\begin{equation}
\rho_{*(p)} = \frac{3\pi}{P^{2}G} \bigg \{\frac{(1 + \sqrt{(1+ \frac{f_2}{f_1})\Delta F_\mathrm{obs}})^{2} - b^2[1 - \sin ^{2}(t_{T}\pi/P)]} {\sin ^{2}(t_{T}\pi/P)}\bigg \}^{3/2}.
\label{sdap}
\end{equation}
While if the planet is hosted by the secondary star the mean stellar density becomes,
\begin{equation}
\rho_{*(s)} = \frac{3\pi}{P^{2}G} \bigg \{\frac{(1 + \sqrt{(1+ \frac{f_1}{f_2})\Delta F_\mathrm{obs}})^{2} - b^2[1 - \sin ^{2}(t_{T}\pi/P)]} {\sin ^{2}(t_{T}\pi/P)}\bigg \}^{3/2}.
\label{sdas}
\end{equation}
Equations \ref{sdap} and \ref{sdas} can be generalized to include additional stellar sources. For example, the stellar density for the primary star in a field which contains n stars is,
\begin{equation}
\rho_{*(p)} = \frac{3\pi}{P^{2}G} \bigg \{\frac{(1 + \sqrt{(\sum_{i=1}^{n} \frac{f_i}{f_1})\Delta F_\mathrm{obs}})^{2} - b^2[1 - \sin ^{2}(t_{T}\pi/P)]} {\sin ^{2}(t_{T}\pi/P)}\bigg \}^{3/2}.
\label{sdag}
\end{equation}
Similarly, the mean stellar density for the secondary in a field of n stars is,
\begin{equation}
\rho_{*(s)} = \frac{3\pi}{P^{2}G} \bigg \{\frac{(1 + \sqrt{(\sum_{i=1}^{n} \frac{f_i}{f_2})\Delta F_\mathrm{obs}})^{2} - b^2[1 - \sin ^{2}(t_{T}\pi/P)]} {\sin ^{2}(t_{T}\pi/P)}\bigg \}^{3/2}.
\label{sdags}
\end{equation}  

\begin{figure*}[htp]
\plotone{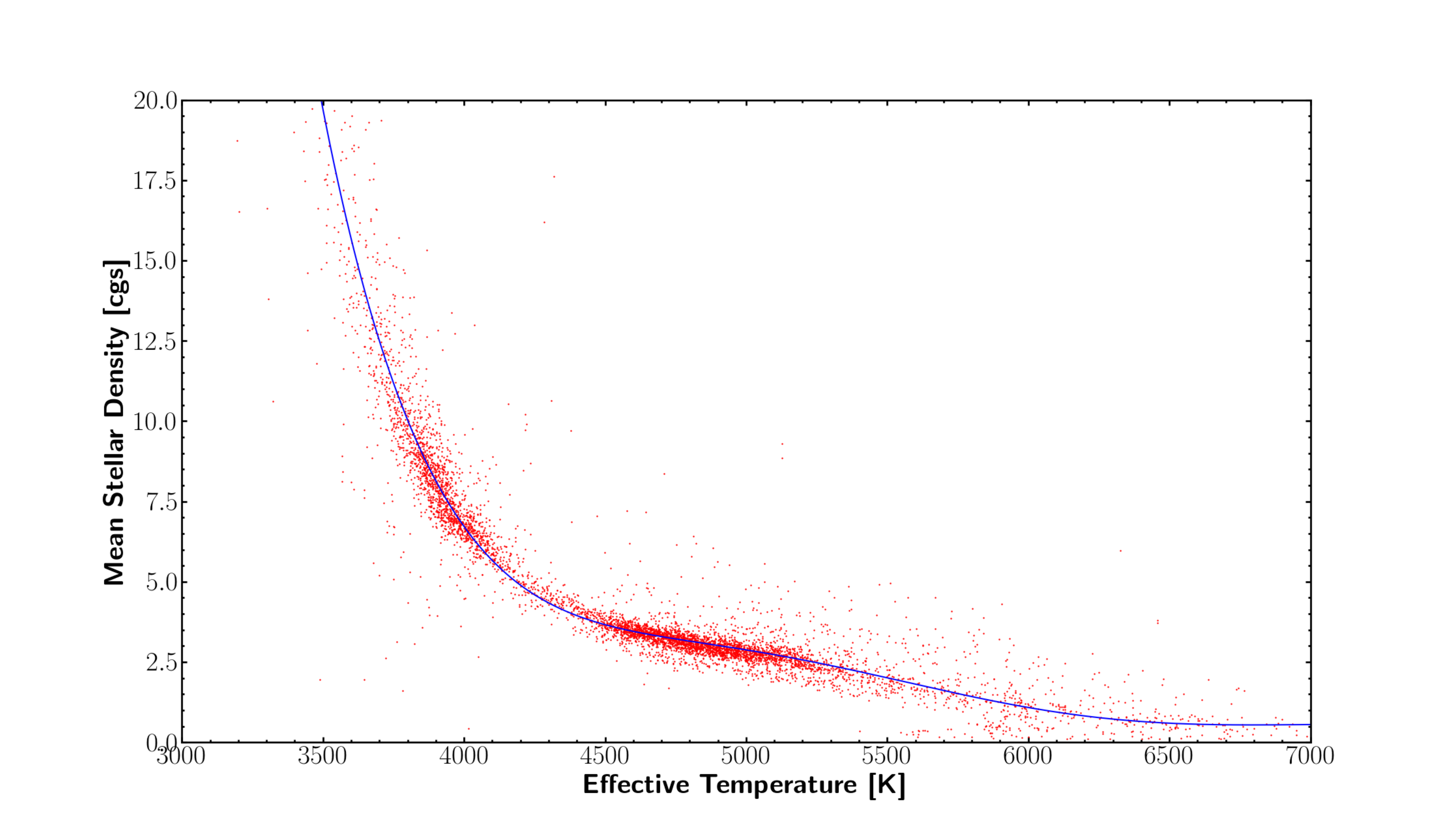}
\caption{Relationship between stellar density and effective temperature. The model, blue line, is a sixth order polynomial and shown here to highlight the relationship between mean stellar density and effective temperature.}
\label{DenTemp}
\end{figure*}

To compare our calculated stellar densities, presented in Table \ref{stellardensity}, to an expected value, we used the same representative sample as described in Section \ref{KpandK} with an additional calculation to derive their mean stellar densities from their stellar radii and masses. The radius was calculated using the Stefan-Boltzman equation, $L = \sigma AT^4$, where A is the surface area, $4\pi R^2$, $T$ is the effective temperature and $L$ is the luminosity. Determining the luminosity required knowing the bolometric magnitudes which we found by first calculating their absolute V magnitude and then converting these magnitudes into bolometric magnitudes using a bolometric correction factor. Once the radii had been determined, we found their stellar masses by using the surface gravity and the derived radius. Finally, the mean stellar density was modeled against effective temperature, see Figure \ref{DenTemp} and Equation \ref{poly6} (sixth order polynomial model), to determine the expected mean stellar densities, which we then used to compare with our target stars.  
\begin{equation}
\begin{split}
\rho_* =   8.752\times10^{-20}(6.942\times10^{-21})T^6_{\mathrm{eff}}\\  
 - 3.271\times10^{-15}(2.370\times10^{-16})T^5_{\mathrm{eff}}\\
 + 5.044\times10^{-11}(3.334\times10^{-12})T^4_{\mathrm{eff}}\\
 - 4.105\times10^{-07}(2.476\times10^{-08})T^3_{\mathrm{eff}}\\  
 + 1.860\times10^{-03}(1.023\times10^{-05})T^2_{\mathrm{eff}}\\
 - 4.451(0.223)T_{\mathrm{eff}}  + 4400.674(200.712) 
 \end{split}
\label{poly6}
\end{equation}
\begin{deluxetable*}{ccccccccccccc}[htp]
  \tabletypesize{\scriptsize}
  \tablecolumns{5}
  \tablecaption{\label{stellardensity} Stellar Density of Target Stars Derived From Transit Fitting}
  \tablehead{
    \colhead{EPIC ID} & \colhead{$\Delta F_\mathrm{obs}$} & \colhead{$\Delta F_\mathrm{obs}$} & \colhead{$b$}  & \colhead{$b$}  & \colhead{$t_T$}  & \colhead{$t_T$}  & \colhead{$Period$} & \colhead{$Period$} & \colhead{$\rho_p$} & \colhead{$\rho_p$} & \colhead{$\rho_s$} & \colhead{$\rho_s$}   \\
    \colhead{System} & \colhead{} & \colhead{Uncertainty}  & \colhead{} & \colhead{Uncertainty} & \colhead{} & \colhead{Uncertainty} & \colhead{} & \colhead{Uncertainty} & \colhead{} & \colhead{Uncertainty} & \colhead{} & \colhead{Uncertainty}  \\ \hline
    \colhead{} & \colhead{} & \colhead{}  & \colhead{} & \colhead{} & \colhead{(seconds)} & \colhead{(seconds)} & \colhead{(seconds)} & \colhead{(seconds)} & \colhead{$gcm^{-3}$} & \colhead{$gcm^{-3}$} & \colhead{$gcm^{-3}$} & \colhead{$gcm^{-3}$} \\ \hline
    \colhead{(1)} & \colhead{(2)} & \colhead{(3)} & \colhead{(4)} & \colhead{(5)}  & \colhead{(6)} & \colhead{(7)} & \colhead{(8)} & \colhead{(9)} & \colhead{(10)}  & \colhead{(11)} & \colhead{(12)}  & \colhead{(13)}  
  }
  \startdata  
201546283 &	0.0023 &	0.00014 &	0.3 &	0.25 &	10152 &	190 &	585043.60 &	5.27 &	2.58663 &	0.59298 &	5.70112 &	1.67882\\
201862715 &	0.0126 &	0.00015 &	0.14 &	0.11 &	9151 &	45 &	229450.56 &	0.04 &	1.99713 &	0.10982 &	2.23983 &	0.20713\\
205148699 &	0.03134 &	0.0003 &	0.08 &	0.08 &	16178 &	86 &	378203.21 &	2.07 &	0.77948 &	0.06664 &	0.81904 &	0.08715\\
206011496 &	0.00028 &	0.00004 &	0.41 &	0.33 &	8009 &	346 &	204695.51 &	6.13 &	1.48168 &	0.70756 &	1.79803 &	0.77331\\
206061524 &	0.00701 &	0.0003 &	0.25 &	0.23 &	8355 &	216 &	507988.80 &	11.23 &	4.78295 &	0.82011 &	6.32927 &	1.26686\\
206192335 &	0.00029 &	0.00007 &	0.54 &	0.36 &	6100 &	510 &	310963.97 &	13.82 &	4.01001 &	3.28912 &	15.27341 &	8.59631\\
210958990 &	0.02051 &	0.00093 &	0.43 &	0.23 &	9297 &	276 &	147079.67 &	1.21 &	1.0631 &	0.2787 &	1.79934 &	0.53495\\
211147528 &	0.00885 &	0.00054 &	0.78 &	0.14 &	9081 &	536 &	202993.00 &	4.84 &	0.58753 &	0.32666 &	16.81957 &	11.67245\\
  \enddata
  \tablecomments{Stellar densities for our primary, column (10) and secondary, column (12) targets stars. Densities were calculated using Equations \ref{sdap} and \ref{sdas}. Column (2) is the observed transit depth, column (4) is the impact parameter, column (6) is the transit duration and column (8) is the period which where all determined through modeling the transit light curve. Also displayed are the uncertainties on each of the tabulated parameters.}
 \end{deluxetable*}

\subsection{Effective Temperature}
\label{effectivetemp}
\begin{figure*}[htp]
\plotone{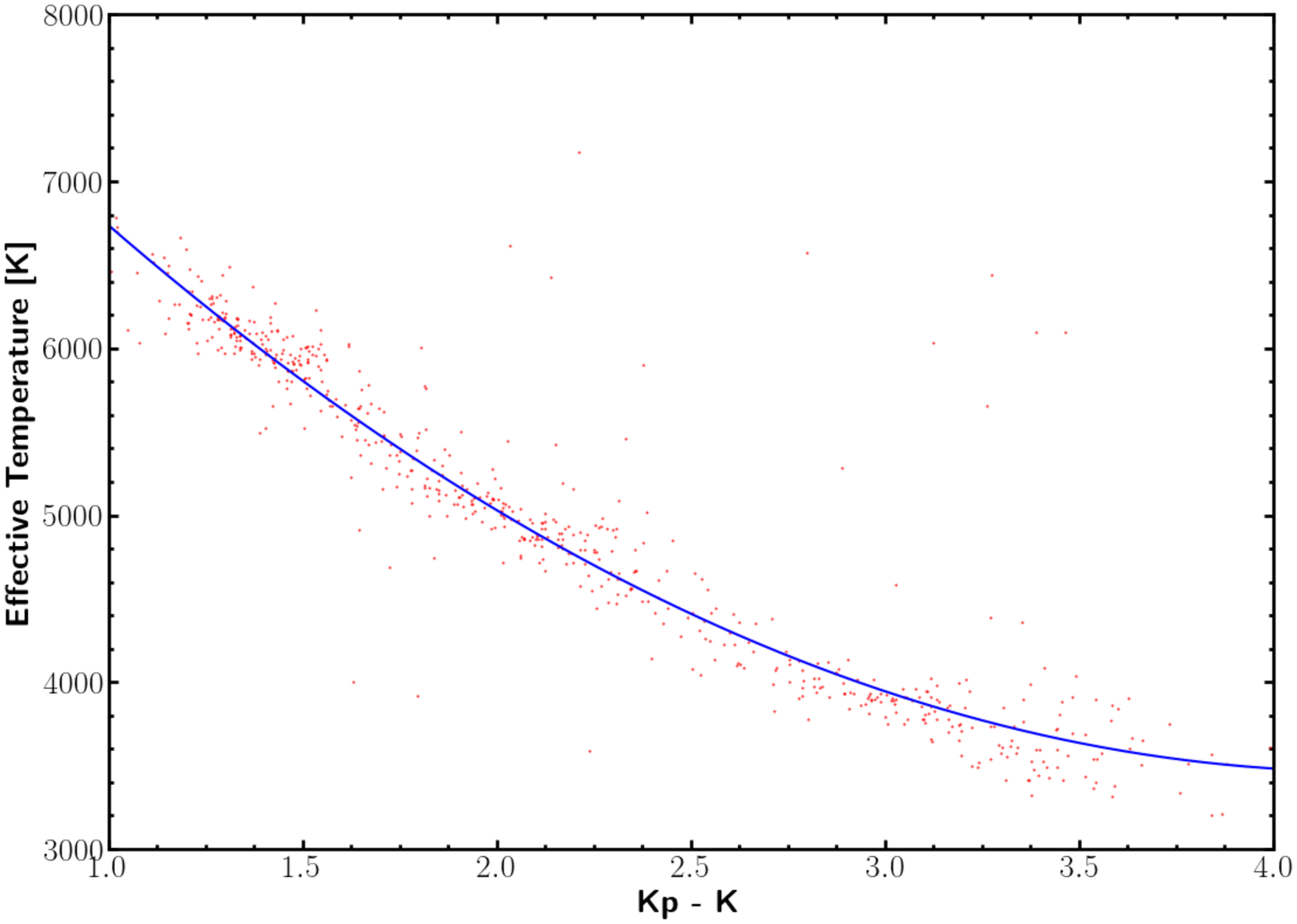}
\caption{Relationship between effective temperature and  $K_p - K$. The model, blue line, is a second order polynomial.}
\label{tempest}
\end{figure*}

In order to estimate the mean stellar densities from Equation \ref{poly6} (Figure \ref{DenTemp}) we need to determine the effective temperature for each of our target stars. We used a combined approach to determine the effective temperature of the our primary and secondary target stars. Firstly, where possible, we used the effective temperature from the second data release of Gaia\footnote{\label{Gaia}This work made use of data from the European Space Agency (ESA) mission {\it Gaia} (\url{https://www.cosmos.esa.int/gaia}), processed by the {\it Gaia} Data Processing and Analysis Consortium (DPAC, \url{https://www.cosmos.esa.int/web/gaia/dpac/consortium}). Funding for the DPAC has been provided by national institutions, in particular the institutions participating in the {\it Gaia} Multilateral Agreement.} and their associated uncertainties. From these data we were able to obtain the effective temperatures for both stars for one of our binary systems, EPIC 201862715. Gaia did not release effective temperatures, or could not resolve the secondary stars in our remaining seven systems. We therefore assumed that the  reported effective temperature from the Gaia database to be the primary effective temperatures for the remaining seven multi-star systems. However, we still needed to find effective temperatures for the remaining seven secondary stars in our targets multi-star systems. To achieve this, we modeled the relationship between the effective temperature and the color $K_p - K$, using the same representative sample as described in Section \ref{KpandK} and is shown in Figure \ref{tempest}.

The data was fitted with a second order polynomial and is shown in Equation \ref{estTeff},
\begin{equation}
\begin{split}
T_{\mathrm{eff}} = 310.85643(24.93115)C^2\\
-  2639.39248(131.63409)C\\ 
+ 9066.15829(164.22003),
\end{split}
\label{estTeff}
\end{equation}
where C is the color index $K_p - K$ and the coefficient uncertainties are reported in the parentheses. The effective temperatures and there uncertainty are tabulated in Table \ref{efftemp}

\begin{deluxetable}{ccccc}[htp]
  \tabletypesize{\scriptsize}
  \tablecolumns{5}
  \tablecaption{\label{efftemp} Effective Temperature of Target Stars}
  \tablehead{
    \colhead{EPIC ID} & \colhead{$T_{\mathrm{eff(p)}}$} & \colhead{$T_{\mathrm{eff(p)}}$} & \colhead{$T_{\mathrm{eff(s)}}$}  & \colhead{$T_{\mathrm{eff(s)}}$}  \\
    \colhead{System} & \colhead{} & \colhead{Uncertainty}  & \colhead{} & \colhead{Uncertainty} \\ \hline
    \colhead{} & \colhead{K} & \colhead{K}  & \colhead{K} & \colhead{K} \\ \hline
    \colhead{(1)} & \colhead{(2)} & \colhead{(3)} & \colhead{(4)} & \colhead{(5)} 
  }
  \startdata  
201546283 &	5105 &	210 &	4354 &	409\\
201862715\textsuperscript{a} & 	5817 &	848 &	5339 &	252\\
205148699 &	5254 &	141 &	4827 &	362\\
206011496 &	5390 &	194 &	4702 &	363\\
206061524 &	4277 &	158 &	4529 &	384\\
206192335 &	5368 &	160 &	4094 &	455\\
210958990 &	4934 &	171 &	4647 &	369\\
211147528 &	6365 &	476 &	4166 &	439\\
  \enddata
  \tablecomments{Effective temperatures for our primary, column (2) and secondary, column (4) targets stars. Effective temperatures for all the secondary stars were derived using Equation \ref{estTeff}, except for EPIC 201862715, which was obtained from the Gaia database, denoted by the superscripted letter a. Effective temperatures for all the primary stars were obtained from the Gaia database. Also displayed are the uncertainties in effective temperature for the primary and secondary.}
 \end{deluxetable}

\section{Determining The Stellar Host} \label{sec:dsh}

In this section, we present the results of our methodology for determining the likely stellar host for our target systems, by comparing the transit derived mean stellar density to the expected means stellar density for each star in our multi-star systems. As outlined in detail in Section \ref{sec:appr} we will use the stellar density and the effective temperature of our target stars to determine which star is likely to be the planets host. This is achieved by investigating which star in each system best fits the mean stellar density--effective temperature relation, as shown in Figure \ref{DenTemp}. 

\begin{figure*}[htp]
\includegraphics[width=20cm,height=15cm]{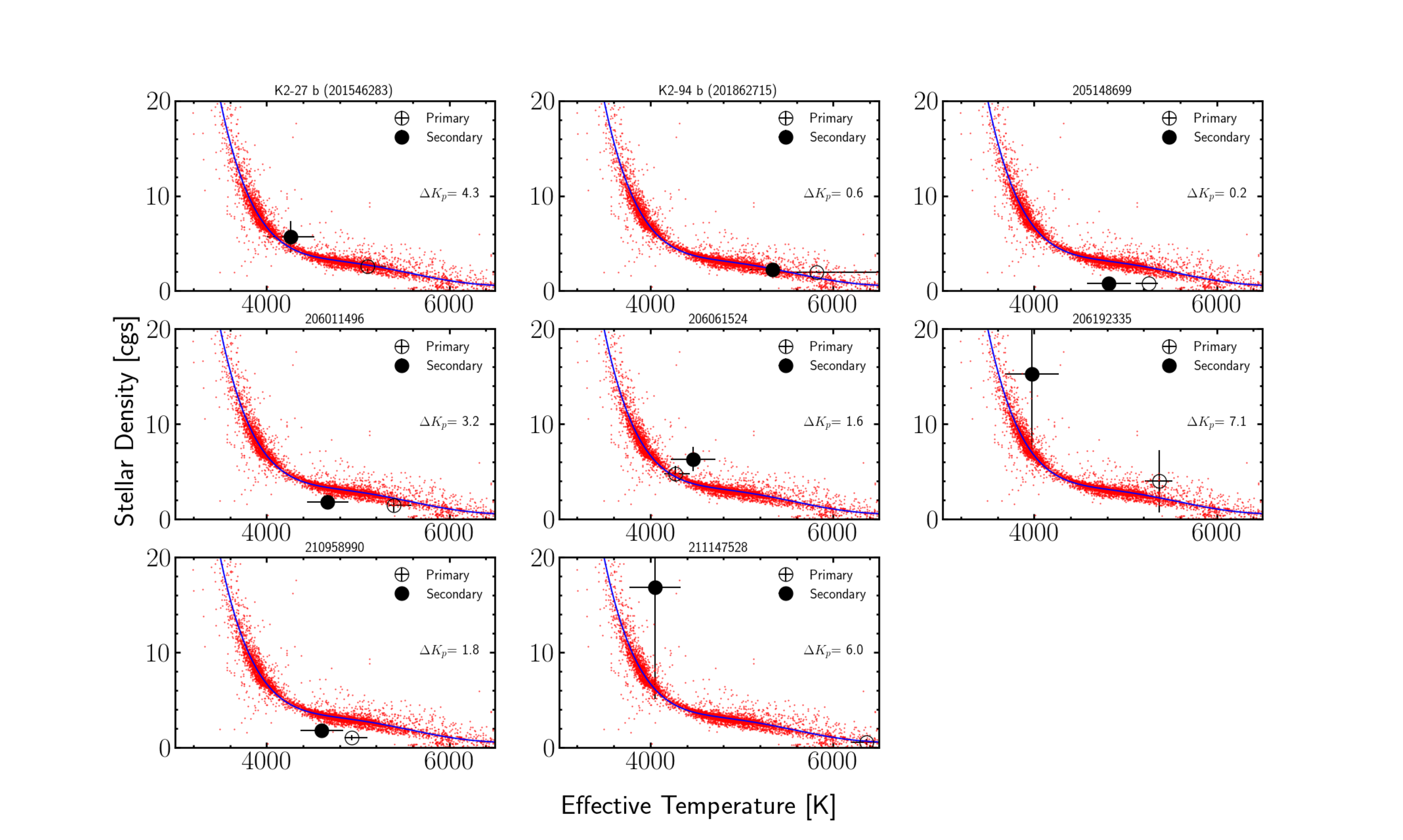}
\caption{Comparison of mean stellar densities derived from transit fitting to expected mean stellar densities for the given stellar effective temperature Also shown is the $\Delta{K_p}$ magnitude for each multi-star system.}
\label{TarDenTemp}
\end{figure*}

Figure \ref{TarDenTemp} reproduces the mean stellar density--effective temperature relationship with our target stars overlaid for each multi-star system. The primary star is represented by an open circle, while the secondary is represented by a closed circle. Two out of the eight target systems have confirmed planets, K2-27b (EPIC 201546283) and K2-97b (EPIC 201862715). These two systems are displayed in the first two panels of Figure \ref{TarDenTemp}, while the remaining systems represent systems with possible planetary candidates and are titled by their EPIC ID.  

It is clear by visual inspection that both the primary and secondary, including their uncertainties, of EPIC 205148699 does not overlap the model and is therefore difficult to make any prediction regarding which star is the planets host. Of the remaining seven multi-star systems systems, six appear to have enough evidence to make a reasonable judgment about which star is the planet hosting star. For four systems, EPIC 206011496, EPIC 206061524, EPIC 206192335 and EPIC 210958990, strong evidence suggests that the planets are hosted by the primary, primary, primary and secondary respectively. The evidence for this comes from the stars that are not suspected to be the planet hosting star because their uncertainties do not overlap the model. While the primary stars of EPIC  201546283 and EPIC 211147528 do not have sufficiently large enough uncertainties not to lie on the model, it is reasonable to assume that the primary stars in these systems are the planet hosting stars. The remaining system, EPIC 201862715, shows no definitive evidence to suggest which star is the planet hosting star, despite the secondary lying directly on the model. Table \ref{sigma} tables the number of standard deviation away the primary and secondary are away from the model.The distance to the model was determined by using the normalized Euclidean distance the primary and secondary were away from the model. The stellar density and effective temperature were normalized by dividing by the median. 
\begin{deluxetable}{ccc}[ht]
  \tabletypesize{\scriptsize}
  \tablecolumns{5}
  \tablecaption{\label{sigma} Distance Away From Model}
  \tablehead{
    \colhead{EPIC ID} & \colhead{$\sigma_{(p)}$} & \colhead{$\sigma_{(s)}$}\\
    \colhead{(1)} & \colhead{(2)} & \colhead{(3)} 
  }
  \startdata  
201546283 &	0.37 &	1.06 \\
201862715 &	0.08 &	0.33 \\
205148699 &	1.13 &	1.73 \\
206011496 &	0.21 &	0.94 \\
206061524 &	0.22 &	2.71 \\
206192335 &	1.19 &	1.29 \\
210958990 &	1.34 &	1.10 \\
211147528 &	0.42 &	5.80 \\
  \enddata
  \tablecomments{Number of standard deviations away from the model.}
 \end{deluxetable}

To investigate the results shown in Figure \ref{TarDenTemp} further, we ran a simulation to test the likelihood that our initial assumptions by visual inspection were reasonable. We ran a Monte Carlo simulation which varied all five parameters in the expressions for the mean stellar density($\Delta F$,$b$,$P$,$t_T$,$f_2/f_1$), Equations \ref{sdap} and \ref{sdas}, and the effective temperature to determine how likely it is for the primary or secondary star to be the planet hosting star. We constructed our simulation by constraining the input parameters to vary between their upper and lower bounds. We ran the simulation under two scenarios: 
\begin{enumerate}
\item Assuming that the distribution of values between the upper and lower bounds of our input parameters and effective temperature were Gaussian distributed.
\item Assuming that any value between the upper and lower bounds was equally likely to be selected. 
\end{enumerate}

The output of each simulation gave a density and an effective temperature for the primary and secondary in each system, which was then used to assess which star lies closest to the model. The results of the simulation are tabulated in Table \ref{MC} and presents the likelihood that the primary or secondary star is the planets host.

\begin{deluxetable}{cccccc}[ht!]
  \tabletypesize{\scriptsize}
  \tablecolumns{5}
  \tablecaption{\label{MC} Planet Hosting Likelihood}
  \tablehead{
    \colhead{EPIC ID} & \colhead{Primary} & \colhead{Secondary} & \colhead{Primary}  & \colhead{Secondary} & \colhead{Likelihood} \\
    \colhead{System} & \colhead{Normal} & \colhead{Normal}  & \colhead{Even} & \colhead{Even} & \colhead{} \\ \hline
    \colhead{} & \colhead{\%} & \colhead{\%}  & \colhead{\%} & \colhead{\%} & \colhead{} \\ \hline
    \colhead{(1)} & \colhead{(2)} & \colhead{(3)} & \colhead{(4)} & \colhead{(5)} & \colhead{(6)} 
  }
  \startdata  
201546283 &	69  &	31 &	70 &	30 & LP\\
201862715 &	49  &	51 &	50 &	50 & U\\
205148699 &	49  &	51 &	40 &	60 & U\\
206011496 &	62  &	38 &	64 &	36 & LP\\
206061524 &	77  &	23 &	75 &	25 & LP\\
206192335 &	95  &	  5 &	95 &	  5 & HLP\\
210958990 &	42  &	58 &	44 &	56 & LS\\
211147528 &	98  &	  2 &	95 &	  5 & HLP\\
  \enddata
  \tablecomments{Results of 10,000 simulations run for both normal and evenly distributed parameter intervals, yielding a likelihood that the planet is hosted by either the primary or the secondary. LP($>$ 55\%) = \lq Likely Primary', U($\approx$ 50\%) = \lq Uncertain', LS($>$ 55\%) = \lq Likely Secondary', HLP($>$ 90\%) = \lq Highly Likely Primary'.}
 \end{deluxetable}

Two systems, EPIC206192335 and EPIC 211147528, the primary star is greater than 90\% more likely to be the planet hosting star and is classified as \textit{highly} likely to be the planet hosting star. The result for EPIC 211147528 confirms the determination of planet hosting star found in Section \ref{KpandK}. Three systems, EPIC 201546283, EPIC 206011496 and EPIC 206061524, the primary star is greater than 55\% more likely to be the planet hosting star and is classified as \textit{most} likely to be the planet hosting star. Given that the visual inspection and the simulation favored the secondary star for EPIC 210958990, the secondary star is likely to be the planet hosting star in this system. Finally, commensurate with our visual inspection results, two systems, EPIC 201862715 and EPIC 205148699, show no conclusive evidence regarding which star is the planet hosting star and therefore remain uncertain.     

\subsection{Factors Influencing the Determination of the Stellar Hosting Star}

\begin{figure*}[ht!]
\plotone{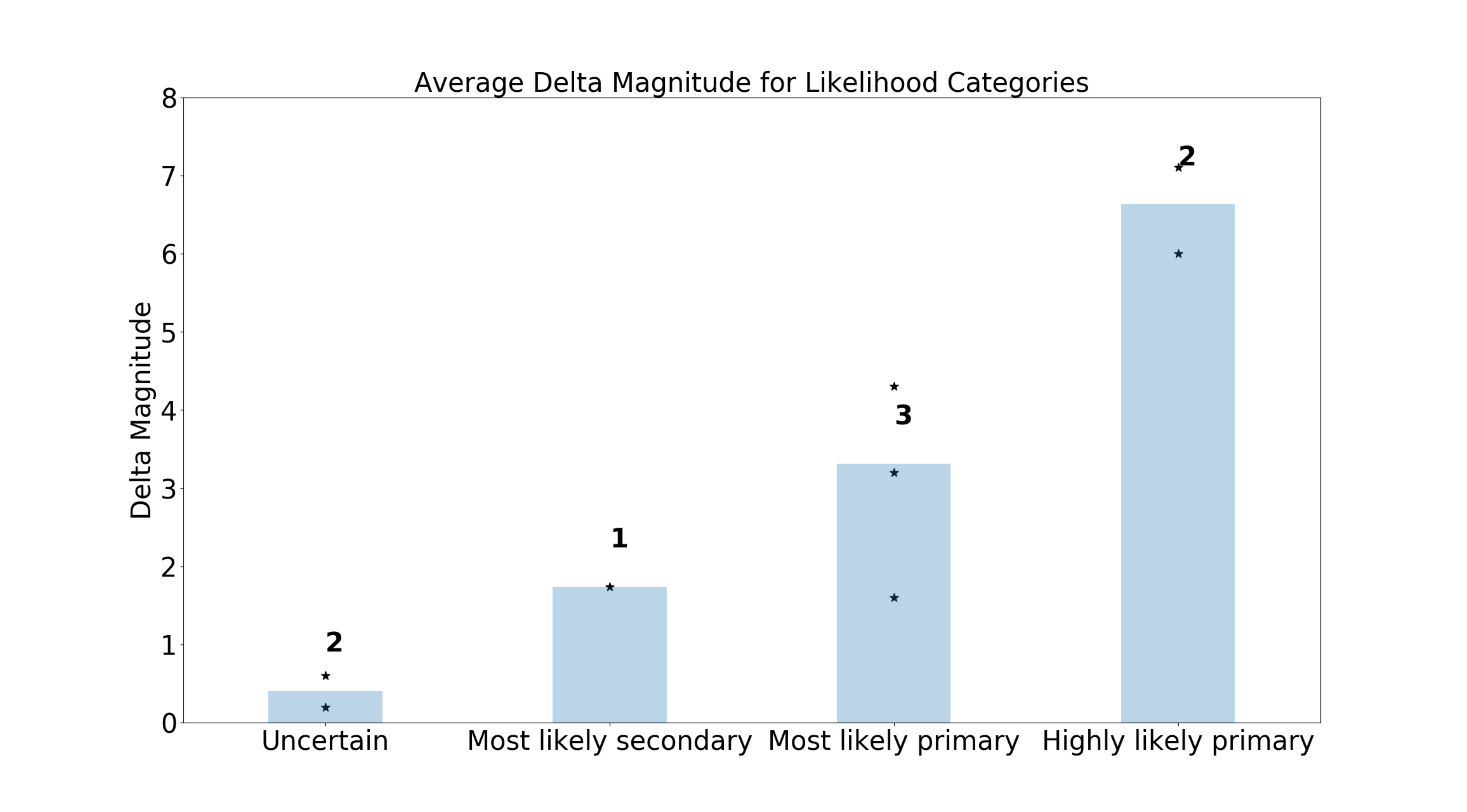}
\caption{The average $K_p$ difference between the primary and secondary stars by likelihood category. The smaller the difference the more uncertainty surrounding the identification of the planet hosting star. The numbers above each bar represent the total stars in each likelihood bin. The star symbols are the $\Delta K_p$ for each multi-star systems.}
\label{ADM}
\end{figure*}

The limitation of our method of determining the planet hosting star is the difference in magnitude between the secondary and primary star. The magnitude difference directly impacts the flux ratio, described in Section \ref{sec:appr}, and hence the determination of the mean stellar density. As the magnitude difference approaches zero, the flux ratio approaches one resulting in the same mean stellar density value for the primary and the secondary stars, thereby making it impossible to identify the planet hosting star. This effect is clearly shown in Figure \ref{ADM}.    

Even if the magnitude difference is high, other factors also impact the level of uncertainty in determining the planet hosting star. Since our methodology is centered around a two dimensional space, mean stellar density and effective temperature, and both of these dimensions depend on other factors, the level of accuracy in their determination directly impacts the level of certainty in determining the planet hosting star. There are seven parameters that can confuse the results: the period; the transit duration; the transit depth; the flux ratio; the impact parameter; and the effective temperature of the primary and the secondary stars. 

\begin{figure*}[ht!]
\includegraphics[width=20cm,height=15cm]{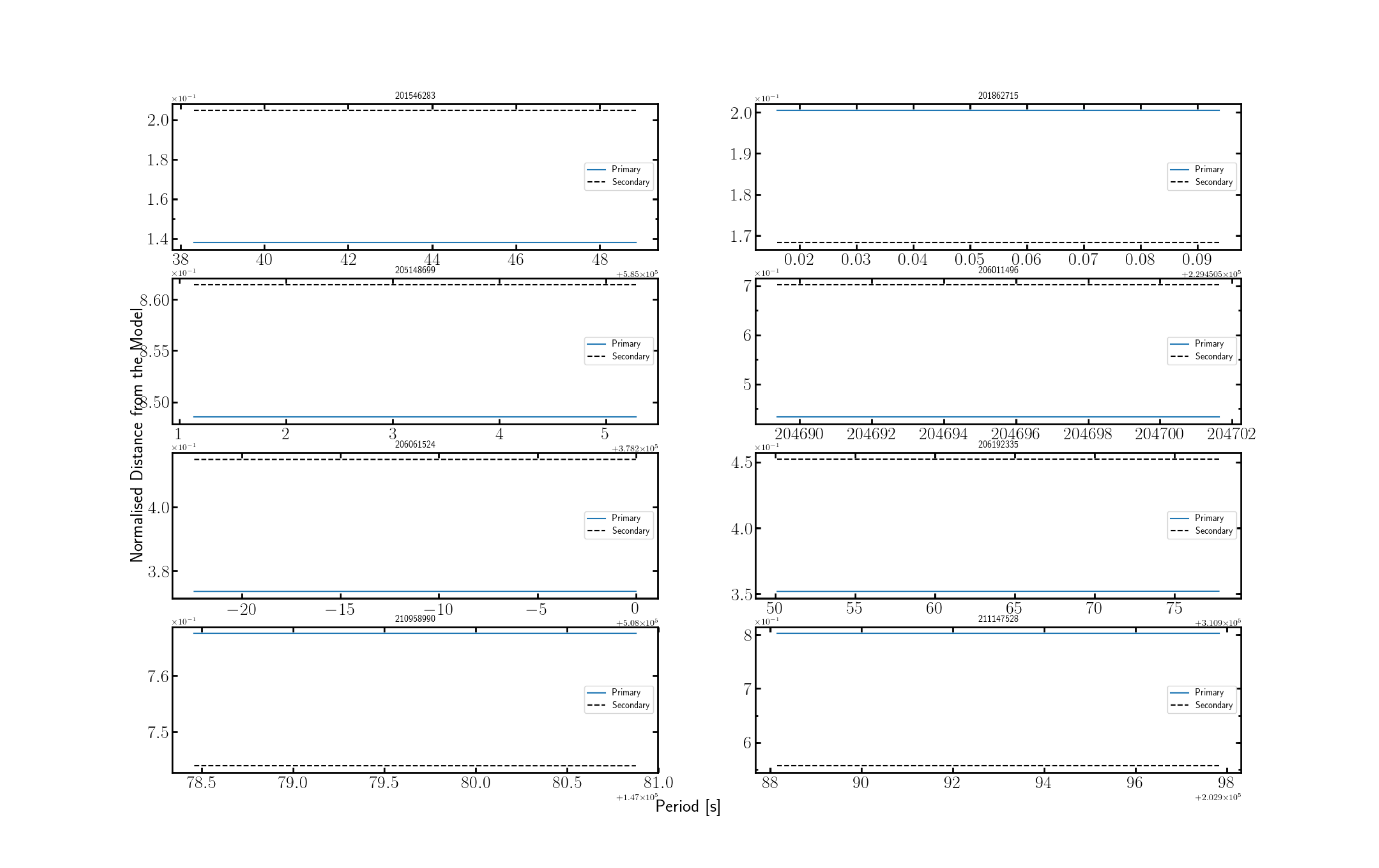}
\caption{Sensitivity analysis showing the effect on the normalized distance to the model form the primary and secondary after taking 1000 evenly spaced samples of the period between the lower and upper error bounds.}
\label{sen1}
\end{figure*}

\begin{figure*}[ht!]
\includegraphics[width=20cm,height=15cm]{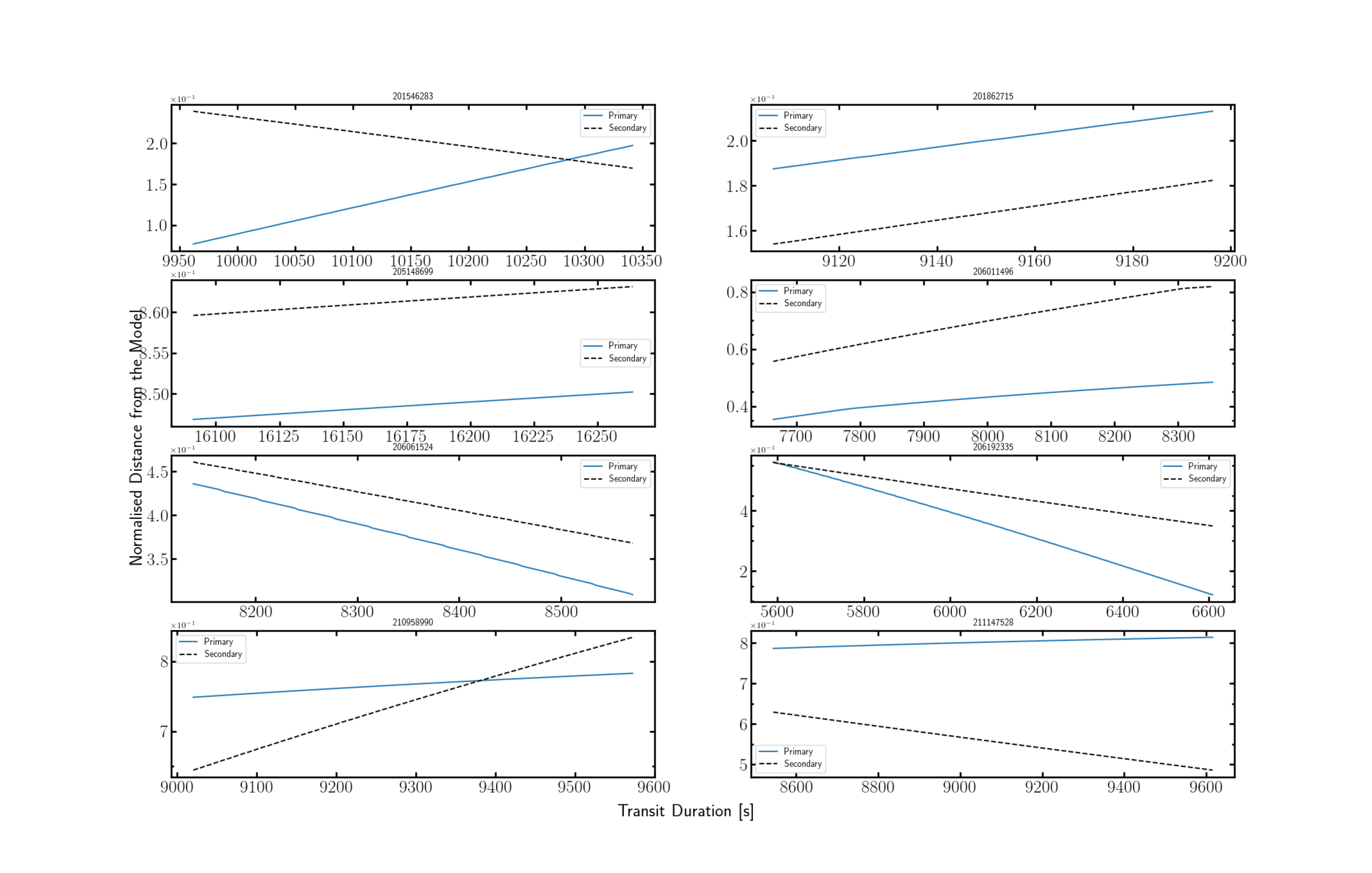}
\caption{Sensitivity analysis showing the effect on the normalized distance to the model form the primary and secondary after taking 1000 evenly spaced samples of the transit duration between the lower and upper error bounds.}
\label{sen2}
\end{figure*}

\begin{figure*}[ht!]
\includegraphics[width=20cm,height=15cm]{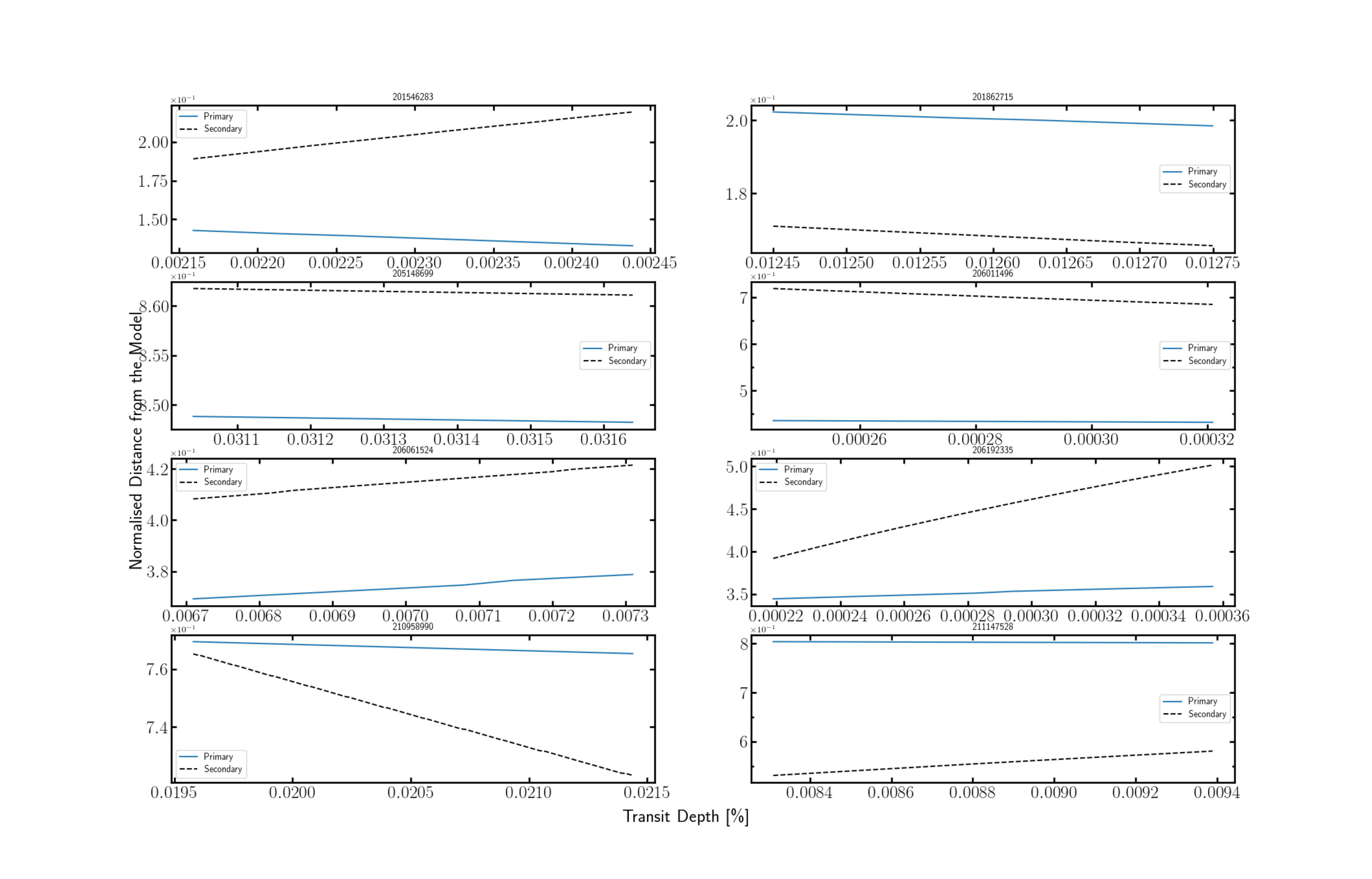}
\caption{Sensitivity analysis showing the effect on the normalized distance to the model form the primary and secondary after taking 1000 evenly spaced samples of the transit depth between the lower and upper error bounds.}
\label{sen3}
\end{figure*}

\begin{figure*}[ht!]
\includegraphics[width=20cm,height=15cm]{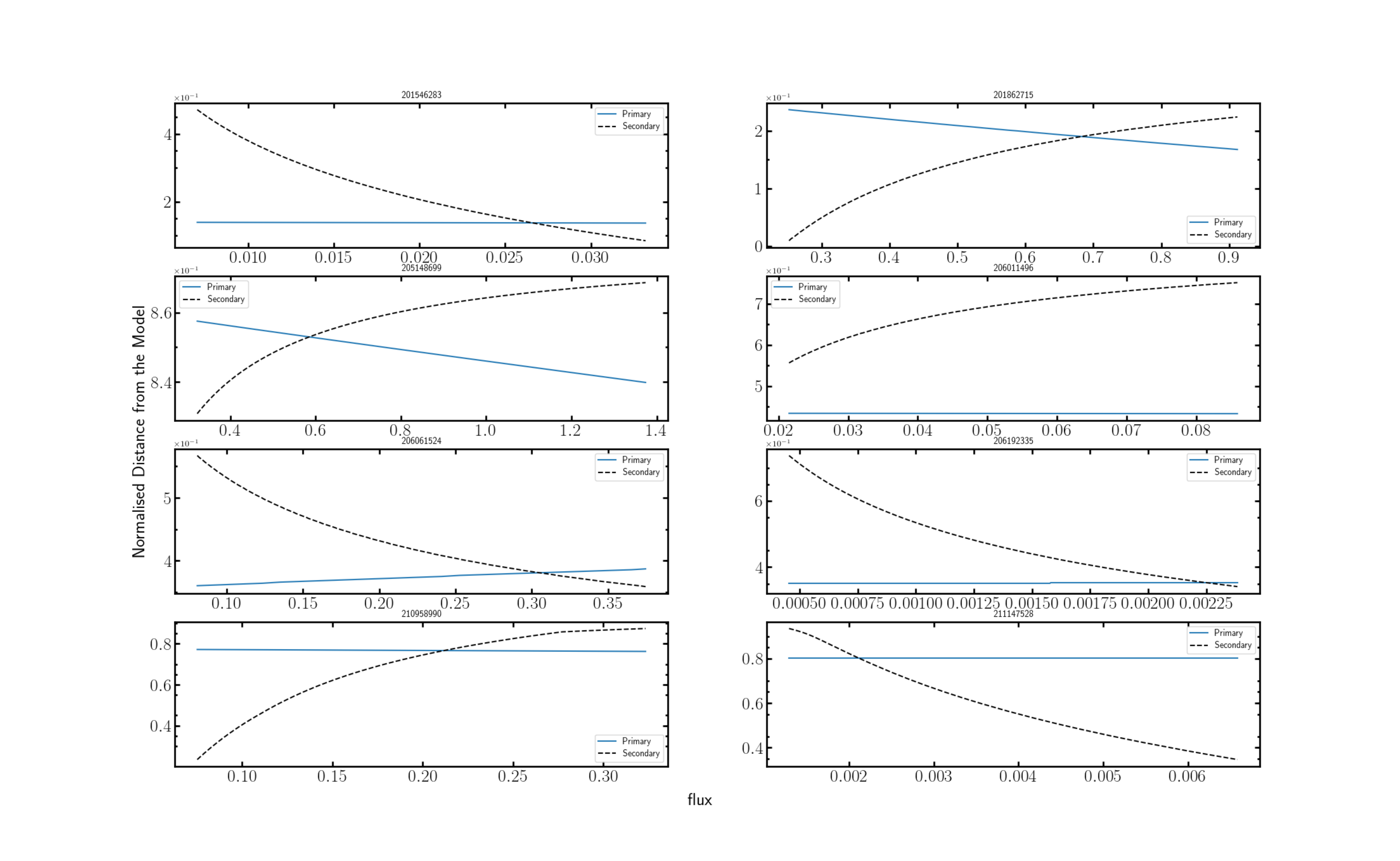}
\caption{Sensitivity analysis showing the effect on the normalized distance to the model form the primary and secondary after taking 1000 evenly spaced samples of the flux ratio between the lower and upper error bounds.}
\label{sen4}
\end{figure*}

\begin{figure*}[ht!]
\includegraphics[width=20cm,height=15cm]{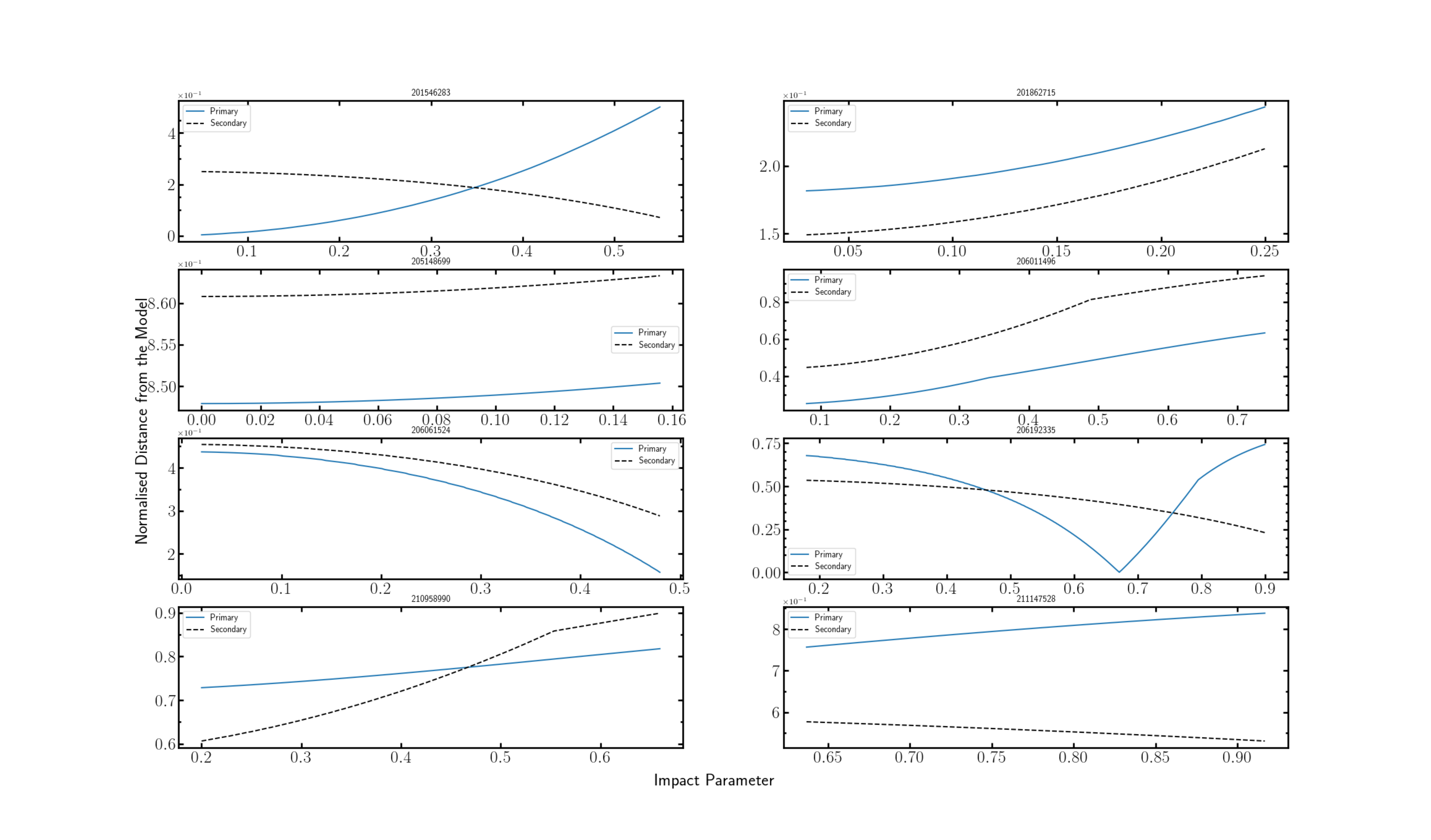}
\caption{Sensitivity analysis showing the effect on the normalized distance to the model form the primary and secondary after taking 1000 evenly spaced samples of the impact parameter between the lower and upper error bounds.}
\label{sen5}
\end{figure*}

\begin{figure*}[ht!]
\includegraphics[width=20cm,height=15cm]{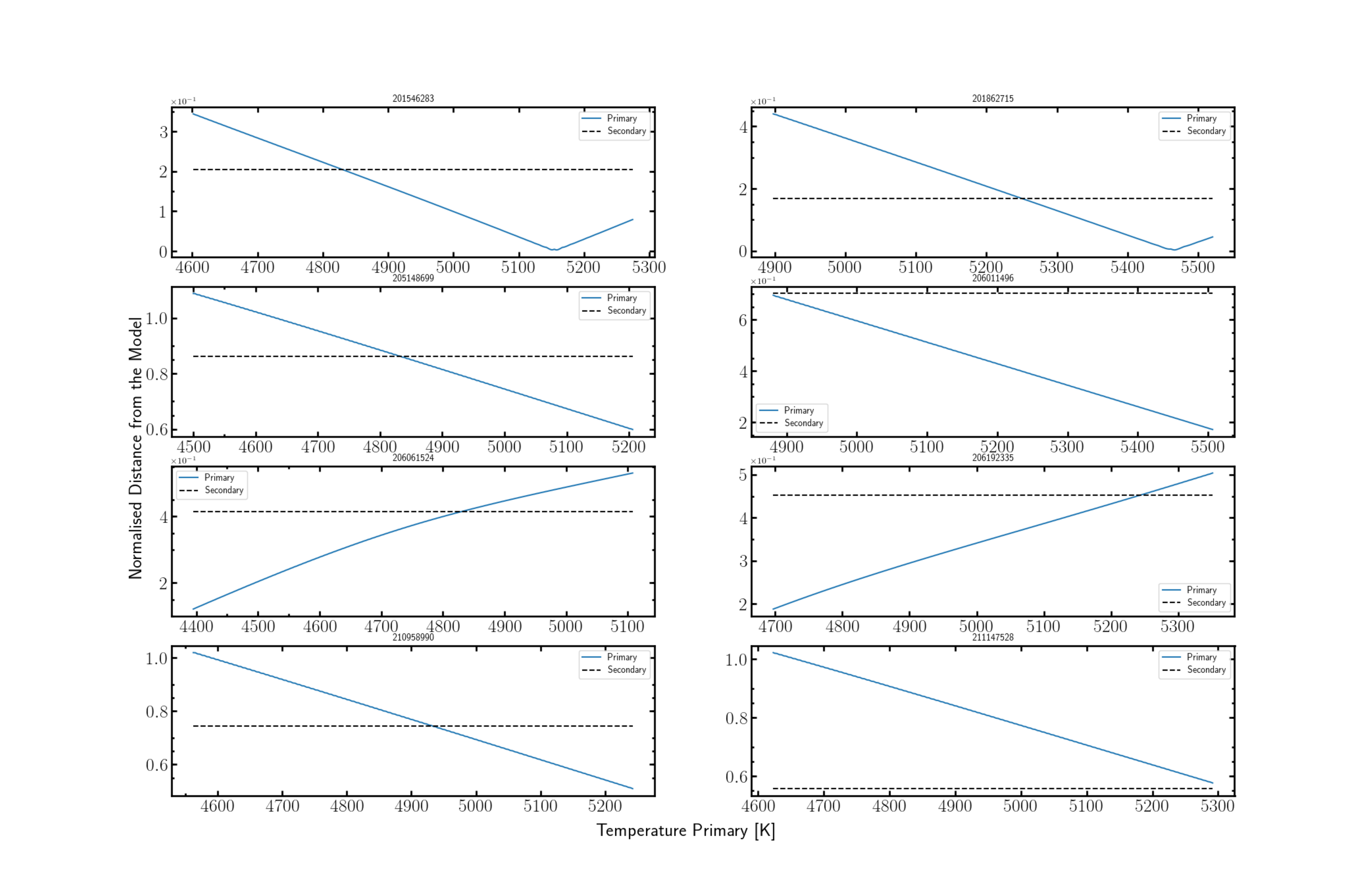}
\caption{Sensitivity analysis showing the effect on the normalized distance to the model form the primary and secondary after taking 1000 evenly spaced samples of the effective temperature of the primary between the lower and upper error bounds.}
\label{sen6}
\end{figure*}

\begin{figure*}[ht!]
\includegraphics[width=20cm,height=15cm]{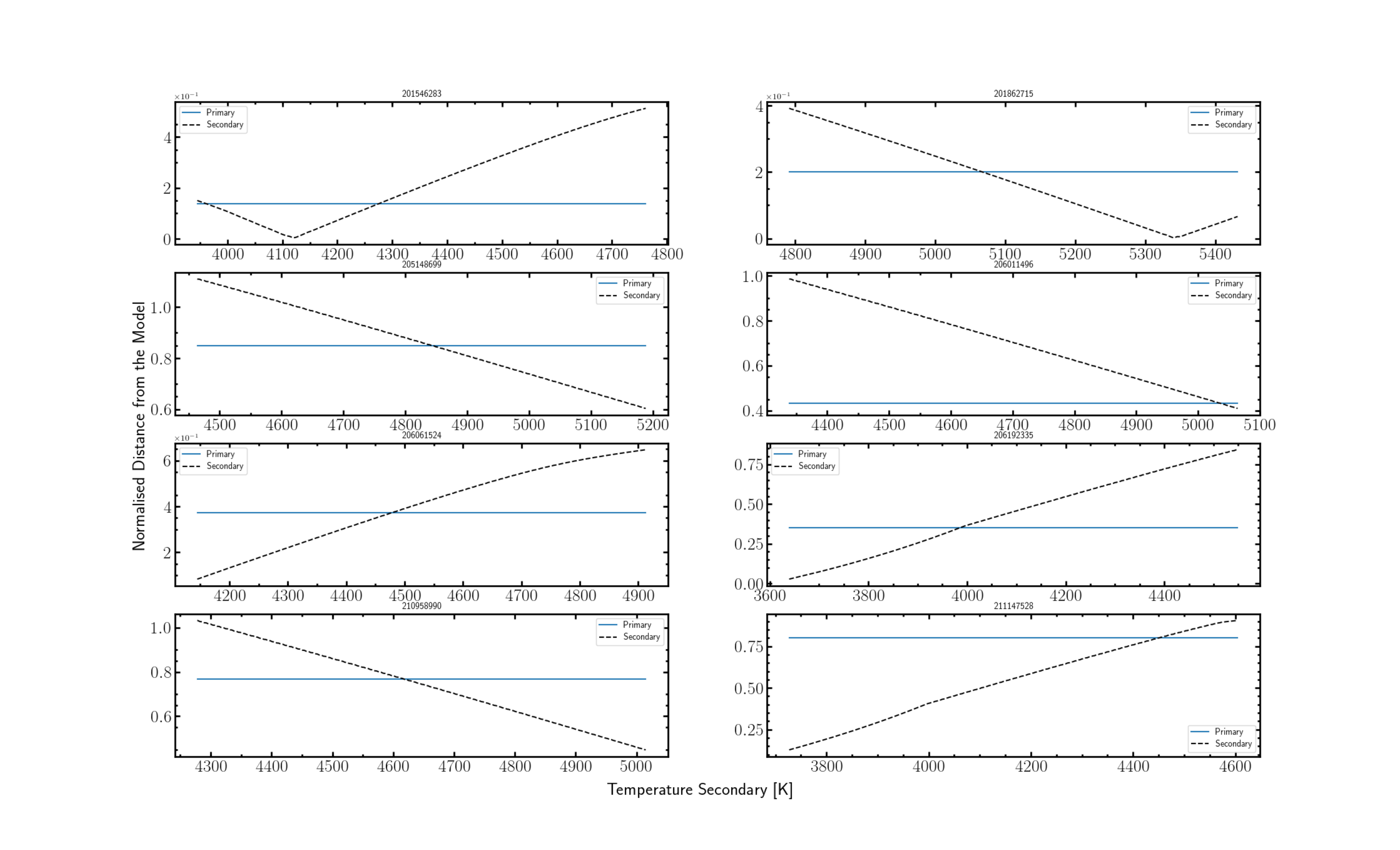}
\caption{Sensitivity analysis showing the effect on the normalized distance to the model form the primary and secondary after taking 1000 evenly spaced samples of the effective temperature of the secondary between the lower and upper error bounds.}
\label{sen7}
\end{figure*}

To examine the impact these parameters had on the determination of the planet hosting star, we ran a sensitivity analysis on each parameter. The normalized distance in density--T\textsubscript{eff} space from the model to the primary and secondary after taking 1000 evenly spaced values from the lower to the upper bound from a parameter, while keeping the remaining six parameters fixed. This was then repeated a further six times, but each time we repeated the process we choose a different parameter to vary. By doing this we were able to see which of the seven parameters created a larger impact on the determination of the planet hosting star. Figures \ref{sen1} to \ref{sen7} present the impact each parameter has on our target systems.

Figure \ref{sen1} clearly shows that varying the period between the upper and lower bounds does not impact the determination of the planet hosting star. This makes sense because the periodicity used in calculating the mean stellar density, Equations \ref{sdap} and \ref{sdas}, for the primary and secondary star are the same (the value of the transit duration, transit depth and impact parameter are also the same for both the primary and the secondary star) and more importantly the width of uncertainty for the period is very small, only deviating at most by 0.004\% from the actual value. This may not be the case if the uncertainty on the period for a system is large.

Figure \ref{sen2} shows the effect the transit duration has on the determination of the planet hosting star. For only two systems out of the eight, EPIC 201546283 and EPIC 210958990, varying the transit duration between its lower and upper bounds (transit duration window) reverses the result. The suspected planet hosting star for EPIC 201546283 (likely primary) and EPIC 206192335 (highly likely secondary) reverses for approximately 15\% and 35\% respectively for the transit duration window. Also, EPIC 206192335 at the lower bound of the transit duration window swaps from the primary to the secondary, but for over 99\% of the interval the primary is revealed as the planet hosting star. Varying the transit duration on the remaining five systems had no effect on the outcome of determining the planet hosting star. The average uncertainty for these five stars is approximately half that of the other three stars indicating that lowering the uncertainty on the transit duration helps in determining the planet hosting star.

Figure \ref{sen3} shows that the observed transit depth (this is not the transit depth of the primary or the secondary, but the transit depth derived from modeling the blended light curve) has very little effect on determining which star is the planet hosting star. The relative uncertainty on the observed transit depth ranges from approximately 1\% for EPIC 205148699 to 24\% for EPIC 206192335 indicating that even with a relatively high uncertainty it doesn't change the outcome. 

Figure \ref{sen4} shows that when varying the flux ratio from its lower to upper bound, all systems except one, EPIC 206011496, have a high degree of uncertainty surrounding which star is the planet hosting star. This can be explained by a high level of uncertainty on \emph{Kepler} bandpass flux ratio itself, with relative errors ranging from approximately 56\% for EPIC 201862715 to 68\% for EPIC 206192335. The uncertainty in the flux ratio, as described in Section \ref{sec:appr}, relies on many factors which can cause the large degree of uncertainty in the flux ratio estimate. In addition, the actual value of the flux ratio itself may be being impacted by additional undetected companion(s).

Figure \ref{sen5} shows that the determination of the planet hosting for three systems; EPIC 201546283, EPIC 206192335 and EPIC 210958990 are sensitive to the value of the impact parameter. The remaining systems reveal that the impact parameter has no impact on the determination of the planet hosting star. The relative uncertainty on the impact parameter is high ranging from approximately 18\% to 100\%, with the three most sensitive systems having relative uncertainties of 83\%, 67\% and 53\% respectively. EPIC 201862715, EPIC 205148699 and EPIC 206061524 have higher uncertainties, but the value of the impact parameter is closer to zero than three sensitive systems, indicating that the relative uncertainty and limb darkening confound the determination of the planet hosting star.     

Figures \ref {sen6} and \ref{sen7} show the effects of varying the temperatures of the primary and the secondary stars between their lower and upper bounds respectively. Both figures clearly show that knowing an accurate temperature of the primary and secondary stars is critical in constraining the determination of the planet hosting star. Most systems exist approximately 50\% of the time being either the primary or the secondary as the planet hosting star. The exception to this is EPIC 206011496 where it appears that the outcome favors the primary as the planet hosting star regardless of whether the primary or the secondary star's temperature is varied. The reason for this can been seen in Figure \ref{TarDenTemp} where the uncertainty in the effective temperature and the density on the primary star is very small and therefore under most cases, will end up being the planet hosting star. 

To fully realize the method outlined in this article for determining the planet hosting star in multi-star systems, the uncertainty on all the parameters which underpin the method must be limited. However, two parameters, the effective temperature of both the primary and secondary stars and the flux ratio are seen as the most influential parameters in determining which star is the planet hosting star. Although, if there is a high degree of certainty in the effective temperature for one of the stars, then this may be sufficient for determining whether or not a star is the planet hosting star. This is because only one star in the system can be the planet hosting star. To control the uncertainty in the flux ratio and effective temperature we recognize the need for additional multiple wavelength high resolution observation to enable the precise determinations of $\Delta K_p$ and $T_{\mathrm{eff}}$ (\cite{2017AJ....153..117H}, \cite{2017AJ....153...71F}). 

\section{Correcting Planetary Radii} \label{sec:cpr}  
 
If we calculated the planetary radius from the observed transit depth for a photometrically blended source,  
\begin{equation}
\Delta F_\mathrm{obs} = R^2_{pb}/R^2_{*b} ,
\label{td}
\end{equation}
where $R_{*b}$ is the pseudo radius of the star and $R_{pb}$ is the pseudo radius of the planet, then it would not only underestimate its true value, but also give an erroneous value because the radius of the stellar source would be incorrect. Equation \ref{td} is valid for our systems because we assume that all observed transits are not grazing (impact parameters are less than 1). We used the term pseudo to highlight that these values are not the true values. To determine the planetary radius for our target planets we first needed to adjust Equation \ref{td} so it corrects the observed transit depth to reflect the transit depth of either the primary or the secondary, depending on which star is the planet hosting star.  If the planet is hosted by the primary star then the radius of the planet can be found by,
\begin{equation}
R^2_p =  \Delta F_{p}R^2_{*p} ,
\label{rp}
\end{equation}
resulting in,
\begin{equation}
R_p =  \sqrt{(1+\frac{f_2}{f_1})\Delta F_\mathrm{obs}R^2_{*p}},
\label{rp1}
\end{equation} 
where $\Delta F{p}$ is the transit depth resulting from the planet orbiting the primary and $R_{*p}$ is the radius of the primary star. Likewise if the planet is hosted by the secondary then radius of the planet can be found by,
\begin{equation}
R_p =  \sqrt{(1+\frac{f_1}{f_2})\Delta F_\mathrm{obs}R^2_{*s}},
\label{rs1}
\end{equation}
where $\Delta F_{s}$ is the transit depth resulting from the planet orbiting the secondary and $R_{*s}$ is the radius of the secondary star.

\begin{figure*}[ht!]
\plotone{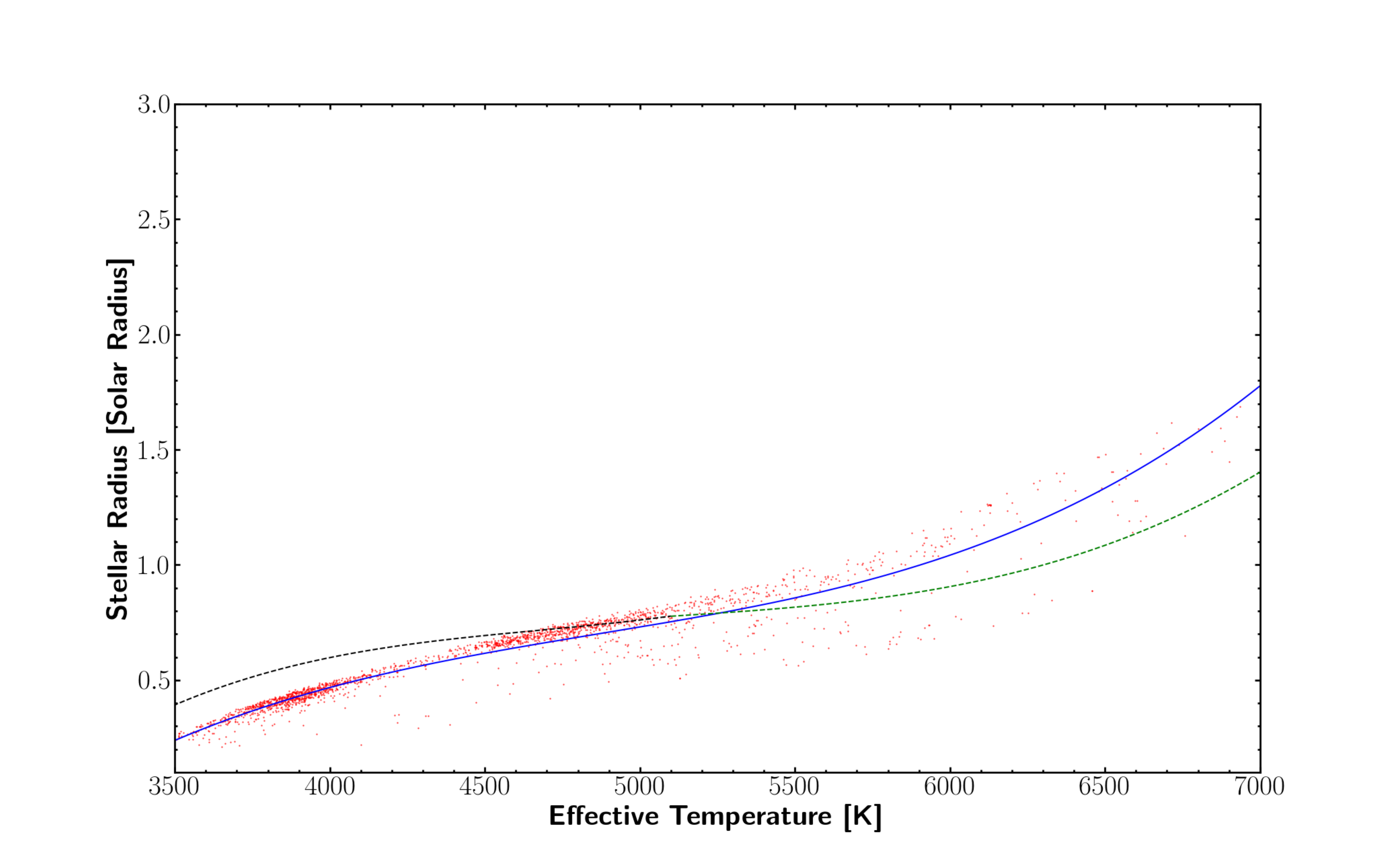}
\caption{The relationship between stellar radius and effective temperature for main sequence dwarf \emph{K2} stars. Also shown, dashed lines, are the two \cite{2012ApJ...757..112B} relations. \cite{2012ApJ...757..112B} split the parameter space into two sections and is represented by the black and green dashed lines. For higher temperatures \cite{2012ApJ...757..112B} constrained the model to pass through through an evolution track that included the Sun.}
\label{RadiusTeff}
\end{figure*}

Since we have the observed transit depth and the flux ratio for our target systems all that remains is to determine the radius of the stars identified as be the most likely to host the planet. The radius of the hosting stars were found by modeling stellar radius against effective temperature. To construct the model we used the same representative sample as described in section \ref{sec:appr} and is shown in Figure \ref{RadiusTeff}. Although the stellar radii and effective temperatures presented in Figure \ref{RadiusTeff} were obtained from the ExoFOP Archive\footref{exofop}, it is noted that these parameters were derived using different methodologies cited in, \cite{2016ApJS..224....2H} and \cite{2016ApJS..226....7C} and therefore may lead to slightly different stellar parameters. The relationship between radius and effective temperature was modeled using a third order polynomial and is presented in Equation \ref{poly5}. The new planetary radii are shown in Table \ref{PlanetRadii}.  We note, this approach was employed by \cite{2012ApJ...757..112B} and the difference between our model and \cite{2012ApJ...757..112B} is shown in Figure \ref{RadiusTeff}  
\begin{equation}
\begin{split}
R_* =   6.321\times10^{-11}(3.556\times10^{-12})T^3_{\mathrm{eff}}\\  
-  9.100\times10^{-07}(5.088\times10^{-08})T^2_{\mathrm{eff}}\\
-  4.652\times10^{-02}(2.397\times10^{-03})T_{\mathrm{eff}}\\
- 7.567(0.372) 
\end{split}
\label{poly5}
\end{equation}

\begin{deluxetable*}{cccccc}[ht!]
\tabletypesize{\scriptsize}
\tablecolumns{5}
\tablecaption{\label{PlanetRadii} Calculated Planetary Radius for Targets}
\tablehead{
\colhead{EPIC ID} & \colhead{Planet Hosting Star} & \colhead{Planet Radius} & \colhead{Planet Radius}  & \colhead{ExoFop Radius} & \colhead{ExoFop Radius} \\
\colhead{System} & \colhead{} & \colhead{}  & \colhead{Uncertainty} & \colhead{} & \colhead{Uncertainty} \\ \hline
\colhead{} & \colhead{} & \colhead{Earth Radii}  & \colhead{Earth Radii} & \colhead{Earth Radii} & \colhead{Earth Radii} \\ \hline
\colhead{(1)} & \colhead{(2)} & \colhead{(3)} & \colhead{(4)} & \colhead{(5)} & \colhead{(6)} 
}
\startdata 
201546283 &	Primary &	4.0  &	0.1  &	4.5  &	0.2 \\
201862715 &	Secondary &	16.4  &	2.9  &	13.9  &	0.3 \\
205148699 &	Secondary &	19.7  &	3.3  &	27  &	0.1 \\
206011496 &	Primary &	1.6  &	0.1  &	1.6  &	0.1 \\
206061524 &	Primary &	5.6  &	0.4  &	6.9  &	0.6 \\
206192335 &	Primary &	1.5  &	0.2  &	1.5  &	0.1 \\
210958990 &	Secondary &	25.0  &	6.5  &	19.1  &	4.2 \\
211147528 &	Primary &	12.8  &	0.4  &	15.8  &	4.6 \\
\enddata
\tablecomments{Planetary radius for planets within our target systems. Column (2) indicates which star is the most probably host. Columns (3) and (4) reveal the calculated radii using either Equation \ref{rp1} or \ref{rs1} and their uncertainties. The uncertainties don't include the uncertainty on the stellar radius. Columns (5) and (6) show the published values from ExoFOP and there uncertainties.}
\end{deluxetable*}

Column 3 in Table \ref{PlanetRadii} present similar radii compared to what is published on the ExoFOP Archive. The uncertainty in the planetary radius is primarily driven by the uncertainty stellar radius. We did not include the uncertainty on the stellar radius and the uncertainties shown in column 4 only reflect the uncertainties associated with the transit depth and flux ratio. In order to include the uncertainty on the stellar radius in the determination of the planetary radius, the predictive power of the model would need to be increased. This would be achieved by including the surface gravity and/or metallicity of the planet hosting star.         
    
\section{Discussion and Conclusion} \label{sec:conc}

Given that the occurrence rate of multi-star systems is expected to be high and the exponential rate of exoplanet detections, there are likely to numerous multi-star systems that contain planetary bodies. These planets are expected to be in circumbinary orbits or in orbits around any or all of the stellar companions within a system. For example, the system EPIC 205703094, not included in our sample, but identified in \cite{2016ApJS..226....7C}, is a binary system that appears to have had three planets detected by \emph{Kepler}. This system was initially included in our sample but rejected because of the lack of reliability in the light curve modeling. Now that we have identified a suitable approach to disentangling photometrically blended transit signals, systems like this one can be analyzed further to determine planet hosting stars, the true radii of the planets and start to gain an appreciation of the overall system architecture. 

Using our our approach to disentangling photometrically blended transit signals we were able to confidently identify the primary star as the planet hosting star in two systems, probably identify four planet hosting stars (3 primary stars and a secondary star) and were uncertain on two out of the eight systems. 

It appears from Figure \ref{ADM} that a strong indicator of the likelihood for identifying which star in a multi-star system is the planet hosting star is the difference in magnitude between the stars in the system. The higher the magnitude difference, the the greater the likelihood of determining the planet hosting star. The difference in magnitude is related by Equation \ref{fluxr} to the flux ratio, which was also shown by Figure \ref{sen4} to strongly impact the determination of the planet hosting star. The relative error on the flux ratio in our study was large ranging from 56\% to 68\% and was in part due to the propagation of errors in converting the photometrically blended \emph{Kepler} magnitudes to \emph{Kepler} magnitudes of the individual stars in each system. To reduce these errors, further ground based observations would need to be carried out. In addition to the follow-up high resolution imaging, follow-up ground based transit observations should be conducted in the same bandpass. Having all data in same bandpass would negate the need to convert between magnitude systems, as was needed in our study, and therefore reduce the error on the flux ratio. 

Furthermore, the accuracy of the effective temperature also played a key role in determining the planet hosting star. To increase the accuracy of the effective temperature, further high resolution ground based follow-up observations need to be undertaken in different band pass filters, or stellar spectra need to be taken to enable the determination of the spectral types of each star in multi-star systems and hence their effective temperatures. This would place less emphasis on model based determinations. 

After the planet hosting star in a multi-star system has been identified, the true radius can then be calculated. We calculated, using either Equation \ref{rp1} and \ref{rs1}, the radii for our eight target planets and presented the results in Table \ref{PlanetRadii}. Our results were hampered by the uncertainty in the stellar radius. To reduce the uncertainty in the stellar radius we would either need to include in our model (Equation \ref{poly5}) the surface gravity and/or metallicity or move away from a model based approach and utilize improved stellar radius estimates via follow-up observations of our planet hosting stars.     

Our approach to disentangling photometrically blended transit signals can be used not only on \emph{Kepler} and \emph{K2} objects, but are easily adapted for any ground or space based transit survey. With data expected in late 2018 or early 2019 from the Transiting Exoplanet Survey Satellite, TESS, our approach is timely. TESS, an all sky survey, which is estimated to collected data on 500,000 plus stars in our local neighborhood, will contain many blended systems, particularly since the pixel area is approximately 25 times that of \emph{Kepler}\footnote{Characteristics of the TESS space telescope: https://heasarc.gsfc.nasa.gov/docs/tess/the-tess-space-telescope.html}. 

In this paper we have clearly outlined an effective methodology for determining the planet hosting star with:
\begin{itemize}
\item{2 out 8 eight multi-star systems (EPIC 206192335 and EPIC 211147528) having a high likelihood the planet hosted by the primary,}
\item{3 out of 8 multi-star systems (EPIC 201546283, EPIC 206011496 and EPIC 206061524) are likely to have the planet hosted by the primary,}
\item{1 out of 8 multi-star systems (EPIC 210958990) is likely to have the planet hosted by the secondary, and}
\item{2 out of 8 multi-star systems (EPIC 201862715 and EPIC 205148699) uncertain which star hosts the planet.}
\end{itemize}
This method is sensitive to the difference in magnitude between the stellar companions in the systems and further investigation will need to be done to determine the minimum magnitude difference threshold for this method to still be effective. In future work, we plan on determining the threshold and further enhancing the statistical robustness of the method by testing it on more than 1000 \emph{Kepler} and \emph{K2} multi-star systems.

\clearpage
\bibliographystyle{aasjournal}
\bibliography{sample62}

\end{document}